\documentclass[conference]{IEEEtran}
\IEEEoverridecommandlockouts
\usepackage{cite}
\usepackage{amsmath,amssymb,amsfonts}
\usepackage{algorithmic}
\usepackage{graphicx}
\usepackage{textcomp}
\usepackage{xcolor}

\def\BibTeX{{\rm B\kern-.05em{\sc i\kern-.025em b}\kern-.08em
    T\kern-.1667em\lower.7ex\hbox{E}\kern-.125emX}}

\begin{document}

\title{Integrated Sensing and Communication for Joint GPS Spoofing and Jamming Detection in Vehicular V2X Networks}

\author{\IEEEauthorblockN{Ali~Krayani, Gabriele~Barabino, Lucio~Marcenaro, Carlo~Regazzoni}
\IEEEauthorblockA{\text{DITEN, University of Genova, Italy} \\
\text{Italian National Inter-University Consortium for Telecommunications (CNIT)}\\
email addresses: \{ali.krayani\}@edu.unige.it , \hspace{0.05cm} \{lucio.marcenaro, carlo.regazzoni\}@unige.it
              }
}

\maketitle

\begin{abstract}
Vehicle-to-everything (V2X) communication is expected to be a prominent component of the sixth generation (6G) to accomplish intelligent transportation systems (ITS). Autonomous vehicles relying only on onboard sensors cannot bypass the limitations of safety and reliability. Thus, integrated sensing and communication is proposed as an effective way to achieve high situational- and self-awareness levels, enabling V2X to perceive the physical world and adjust its behaviour to emergencies. Secure navigation through the Global Positioning System (GPS) is essential in ITS for safe operation. Nevertheless, due to the lack of encryption and authentication mechanisms of civil GPS receivers, spoofers can easily replicate satellite signals by launching GPS spoofing attacks to deceive the vehicle and manipulate navigation data. In addition, due to its shared nature, V2X links are prone to jamming attacks which might endanger vehicular safety. This paper proposes a method to jointly detect GPS spoofing and jamming attacks in a V2X network. Simulation results demonstrate that the proposed method can detect spoofers and jammers with high detection probabilities.
\end{abstract}


\section{Introduction}
In recent years, vehicular technology has attracted significant attention from the automotive and telecommunication industries, leading to the emergence of vehicle-to-everything (V2X) communications for improving road safety, traffic management services and driving comfort.
V2X supported by the sixth generation (6G) is envisioned to be a key enabler of future connected autonomous vehicles \cite{9779322}. Although its transformative benefits for leveraging intelligent transportation systems, V2X still face several technical issues mainly related to performance and security.

The integration of sensing and communication (ISAC) has emerged very recently as a revolutionary element of 6G that could potentially help enabling adaptive learning and intelligent decision-making in future V2X applications.
The combination of sensing and communication allows vehicles to perceive their surroundings better, predict manoeuvres from nearby users and make intelligent decisions, thus paving the way toward a safer transportation system \cite{9665433}.
Modernized vehicles are augmented with various types of sensors divided into exteroceptive to observe their surrounding environment and proprioceptive to observe their internal states.
The former like GPS, Lidar, and Cameras are conveyed to improve situational awareness, while latter sensors, such as steering, pedal, and wheel speed, convey to improve self-awareness. 

While sensing the environment, vehicles can exchange messages that assist in improving situational- and self-awareness and in coordinating maneuvers with other vehicles.
Those messages like the basic safety (BSMs) and cooperative awareness messages (CAMs) are composed of transmitting vehicle's states such as position and velocity and other vehicles' states in the vicinity. Vehicles might use their sensors, such as cameras and Lidar, to detect road users (e.g., pedestrians), which can be communicated with other road users via the V2X messages to improve the overall performance. However, V2X communication links carrying those messages are inherently vulnerable to malicious attacks due to the open and shared nature of the wireless spectrum among vehicles and other cellular users \cite{8336901}. For instance, a jammer in the vicinity might alter the information to be communicated to nearby vehicles/users or can intentionally disrupt communication between a platoon of vehicles making the legitimate signals unrecognizable for on-board units (OBUs) and/or road side units (RSUs) that endanger vehicular safety 
\cite{8553649}.

In addition, the integrity of GPS signals and the correct acquisition of navigation data to compute position, velocity and time information is critical in V2X applications for their safe operation. However, since civil GPS receivers rely on unencrypted satellite signals, spoofers can easily replicate them by deceiving the GPS receiver to compute falsified positions \cite{9226611}.
Also, the long distance between satellites and terrestrial GPS receivers leads to an extremely weak signal that can be easily drowned out by a spoofer. 
Thus, GPS sensors' vulnerability to spoofing attacks poses a severe threat that might be causing vehicles to be out of control or even hijacked and endanger human life \cite{9881548}.
Therefore, GPS spoofing attacks and jamming interference needs to be controlled and detected in real-time to reach secured vehicular communications allowing vehicles to securely talk to each other and interact with the infrastructure (e.g., roadside terminals, base stations) \cite{9860410}.

Existing methods for GPS spoofing detection include GPS signal analysis methods and GPS message encryption methods \cite{9845684}. However, the former requires the ground truth source during the detection process, which is not always possible to collect. In contrast, the latter involves support from a secured infrastructure and advanced computing resources on GPS receivers, which hinders their adoption in V2X applications. On the other hand, existing methods for jammer detection in vehicular networks are based on analysing the packet drop rate as in \cite{9484071}, making it difficult to detect an advanced jammer manipulating the legitimate signal instead of disrupting it.
In this work, we propose a method to jointly detect GPS spoofing and jamming attacks in the V2X network. A coupled generalized dynamic Bayesian network (C-GDBN) is employed to learn the interaction between RF signals received by the RSU from multiple vehicles and their corresponding trajectories. This integration of vehicles' positional information with vehicle-to-infrastructure (V2I) communications allows semantic learning while mapping RF signals with vehicles' trajectories and enables the RSU to jointly predict the RF signals it expects to receive from the vehicles from which it can anticipate the expected trajectories.

The main contributions of this paper can be summarized as follows: \textit{i)} A joint GPS spoofing and jamming detection method is proposed for the V2X scenario, which is based on learning a generative interactive model as the C-GDBN. Such a model encodes the cross-correlation between the RF signals transmitted by multiple vehicles and their trajectories, where their semantic meaning is coupled stochastically at a high abstraction level. \textit{ii)} A cognitive RSU equipped with the acquired C-GDBN can predict and estimate vehicle positions based on real-time RF signals. This allows RSU to evaluate whether both RF signals and vehicles' trajectories are evolving according to the dynamic rules encoded in the C-GDBN and, consequently, to identify the cause (i.e., a jammer attacking the V2I or a spoofer attacking the satellite link) of the abnormal behaviour that occurred in the V2X environment. \textit{iii)} Extensive simulation results demonstrate that the proposed method accurately estimates the vehicles' trajectories from the predicted RF signals, effectively detect any abnormal behaviour and identify the type of abnormality occurring with high detection probabilities.
To our best knowledge, this is the first work that studies the joint detection of jamming and spoofing in V2X systems.

\section{System model and problem formulation}
\begin{figure}[t!]
    \centering
    \includegraphics[height=5.3cm]{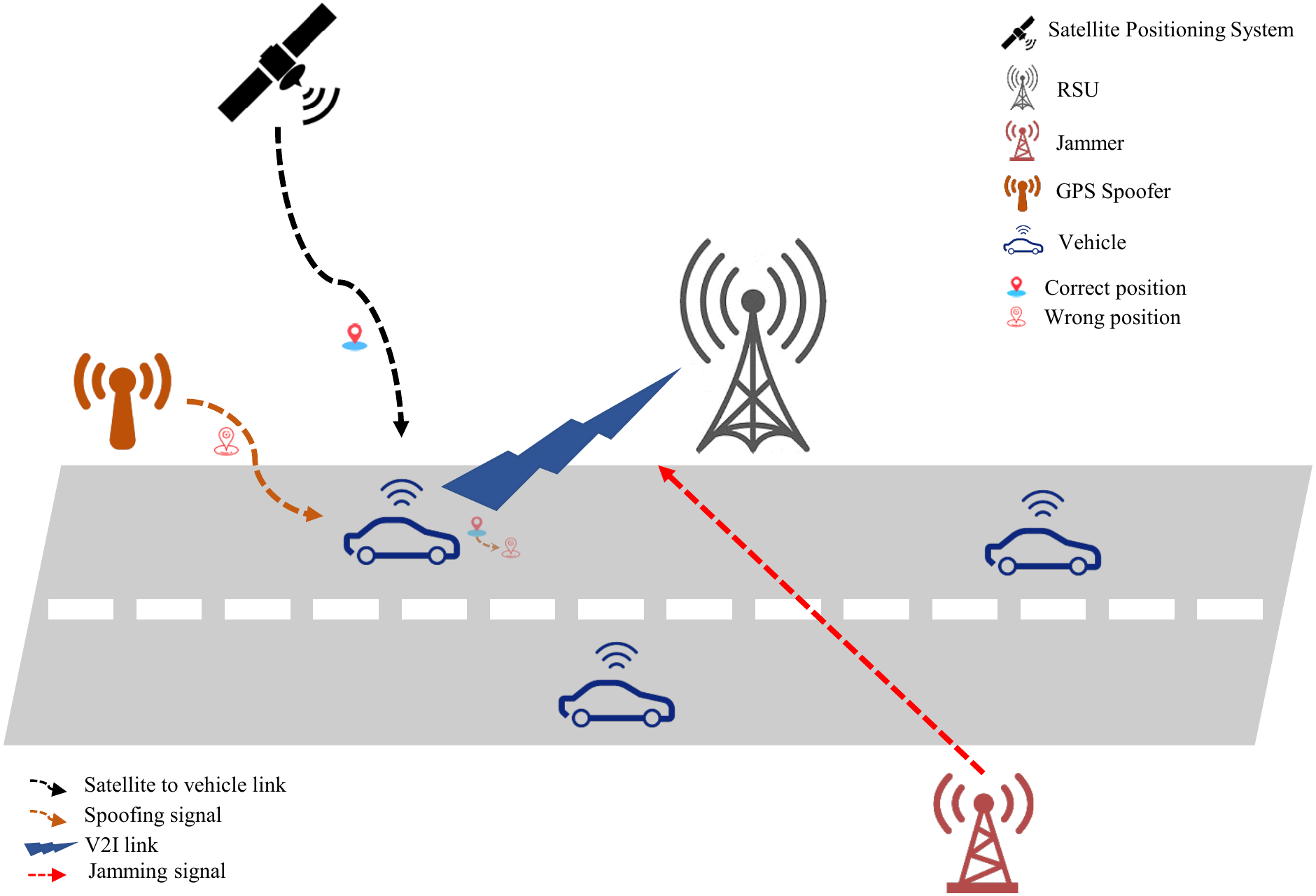}
    \caption{An illustration of the system model.}
    \label{fig_SystemModel}
\end{figure}

The system model depicted in Fig.~\ref{fig_SystemModel}, includes a single cell vehicular network consisting of a road side unit (RSU) located at $\mathrm{p}_{R}=[{x}_{R},{y}_{R}]$, a road side jammer (RSJ) located at $\mathrm{p}_{J}=[{x}_{J},{y}_{J}]$, a road side spoofer (RSS) located at $\mathrm{p}_{s}=[{x}_{s},{y}_{s}]$ and $N$ vehicles moving along multi-lane road in an urban area. The time-varying positions of the $n$-th vehicle is given by $\mathrm{p}_{n,t}=[{x}_{n,t},{y}_{n,t}]$ where $n \in N$. Among the $K$ orthogonal subchannels available for the Vehicle-to-Infrastructure (V2I) communications, RSU assigns one V2I link to each vehicle. Each vehicle exchanges messages composed of the vehicle's state (i.e., position and velocity) with RSU through the $k$-th V2I link by transmitting a signal $\textrm{x}_{t,k}$ carrying those messages at each time instant $t$ where $k \in K$. We consider a reactive RSJ that aims to attack the V2I link by injecting intentional interference to the communication link between vehicles and RSU to alter the transmitted signals by the vehicles. In contrast, the RSS purposes to mislead the vehicles by spoofing the GPS signal and so registering wrong GPS positions. RSU aims to detect both the spoofer on the satellite link and the jammer on multiple V2I links in order to take effective actions and protect the vehicular network. 
%
The joint GPS spoofing and jamming detection problem can be formulated as the following ternary hypothesis test:
\begin{equation}
    \begin{cases}
        \mathcal{H}_{0}: \mathrm{z}_{t,k} = \mathrm{g}_{t,k}^{nR} \mathrm{x}_{t,k} + \mathrm{v}_{t,k}, \\
        \mathcal{H}_{1}: \mathrm{z}_{t,k} = \mathrm{g}_{t,k}^{nR} \mathrm{x}_{t,k} + \mathrm{g}_{t,k}^{JR} \mathrm{x}_{t,k}^{j} + \mathrm{v}_{t,k}, \\
        \mathcal{H}_{2}: \mathrm{z}_{t,k} = \mathrm{g}_{t,k}^{nR} \mathrm{x}_{t,k}^{*} + \mathrm{v}_{t,k},
    \end{cases}
\end{equation}
where $\mathcal{H}_{0}$, $\mathcal{H}_{1}$ and $\mathcal{H}_{2}$ denote three hypotheses corresponding to the absence of both jammer and spoofer, the presence of the jammer, and the presence of the spoofer, respectively. $\textrm{z}_{t,k}$ is the received signal at the RSU at $t$ over the $k$-th V2I link, $\textrm{g}_{t,k}^{nR}$ is the channel power gain from vehicle $n$ to the RSU formulated as: $\textrm{g}_{t,k}^{nR} = \alpha_{t,k}^{nR} \mathrm{h}_{t,k}^{nR}$, where $\alpha_{t,k}^{nR}$ is the large-scale fading including path-loss and shadowing modeled as \cite{8723178}: $\alpha_{t,k}^{nR}=G\beta d_{t,nR}^{-\gamma}$.
$G$ is the pathloss constant, $\beta$ is a log normal shadow fading random variable, $d_{t,nR}=\sqrt{({x}_{n,t}-x_{R})^{2}+({y}_{n,t}-y_{R})^{2}}$ is the distance between the $n$-th vehicle and the RSU. $\gamma$ is the power decay exponent and
$\mathrm{h}_{t,k}$ is the small-scale fading component distributed according to $\mathcal{CN}(0,1)$. In addition, $\mathrm{x}_{t,k}$ is the desired signal transmitted by the $n$-th vehicle, and $\mathrm{v}_{t,k}$ is an additive white Gaussian noise with variance $\sigma_{n}^{2}$. $\mathrm{x}_{t,k}^{J}$ is the jamming signal, $\mathrm{x}_{t,k}^{*}$ is the spoofed signal (i.e., the signal that carries the bits related to the wrong GPS positions), $\mathrm{g}_{t,k}^{JR} = \alpha_{t,k}^{JR} \mathrm{h}_{t,k}^{JR}$ is the channel power gain from RSJ to RSU where $\alpha_{t,k}^{JR}=G\beta d_{t,JR}^{-\gamma}$ such that $d_{t,JR}=\sqrt{({x}_{J}-x_{R})^{2}+({y}_{J}-y_{R})^{2}}$.
We assume that the channel state information (CSI) of V2I links is known and can be estimated at the RSU as in \cite{8345717}. 
The RSU is equipped with an RF antenna which can track the vehicles' trajectories after decoding the received RF signals. RSU aims to learn the interaction between the RF signals received from multiple vehicles and their corresponding trajectories.

\section{Proposed method for joint detection of GPS spoofing and jamming}

\subsection{Environment Representation}
The RSU is receiving RF signals from each vehicle and tracking its trajectory (which we refer to as GPS signal) by decoding and demodulating the received RF signals. 
The Generalized state-space model describing the $i$-th signal evolvement at multiple levels embodies the following equations: 
\begin{equation} \label{eq_discreteLevel}
    \mathrm{\Tilde{S}_{t}}^{(i)} = \mathrm{f}(\mathrm{\Tilde{S}_{t-1}}^{(i)}) + \mathrm{\tilde{w}}_{t},
\end{equation}
\begin{equation} \label{eq_continuousLevel}
    \mathrm{\Tilde{X}_{t}}^{(i)} = \mathrm{A} \mathrm{\Tilde{X}_{t-1}}^{(i)} + \mathrm{B} \mathrm{U}_{\mathrm{\Tilde{S}_{t}}^{(i)}} + \mathrm{\tilde{w}}_{t},
\end{equation}
\begin{equation} \label{eq_observationLevel}
    \mathrm{\Tilde{Z}_{t}}^{(i)} = \mathrm{H} \mathrm{\Tilde{X}_{t}}^{(i)} + \mathrm{\tilde{v}}_{t},
\end{equation}
where $i \in \{$RF, GPS$\}$ indicates the type of signal received by the RSU. The transition system model defined in \eqref{eq_discreteLevel} explains the evolution of the discrete random variables $\mathrm{\Tilde{S}_{t}}^{(i)}$ representing the clusters of the RF (or GPS) signal dynamics, $\mathrm{f}(.)$ is a non linear function of its argument and the additive term $\mathrm{\tilde{w}}_{t}$ denotes the process noise. The dynamic model defined in \eqref{eq_continuousLevel} explains the RF signal dynamics evolution or the motion dynamics evolution of the $n$-th vehicle, where $\mathrm{\Tilde{X}_{t}}^{(i)}$ are hidden continuous variables generating sensory signals, $\mathrm{A} \in \mathbb{R}^{2d}$ and $\mathrm{B} \in \mathbb{R}^{2d}$ are the dynamic and control matrices, respectively, and $\mathrm{U}_{\mathrm{\Tilde{S}_{t}}^{(i)}}$ is the control vector representing the dynamic rules of how the signals evolve with time. The measurement model defined in \eqref{eq_observationLevel} describes dependence of the sensory signals $\mathrm{\Tilde{Z}_{t}}^{(i)}$ on the hidden states $\mathrm{\Tilde{X}_{t}}^{(i)}$ that is parametrized by the measurement matrix $\mathrm{B} \in \mathbb{R}^{2d}$ where $d$ stands for the data dimensionality and $\mathrm{\tilde{v}}_{t}$ is a random noise. 

\subsection{Learning GDBN}
\begin{figure}[t!]
    \begin{center}
        \begin{minipage}[b]{.40\linewidth}
        \centering
           \includegraphics[width=2.5cm]{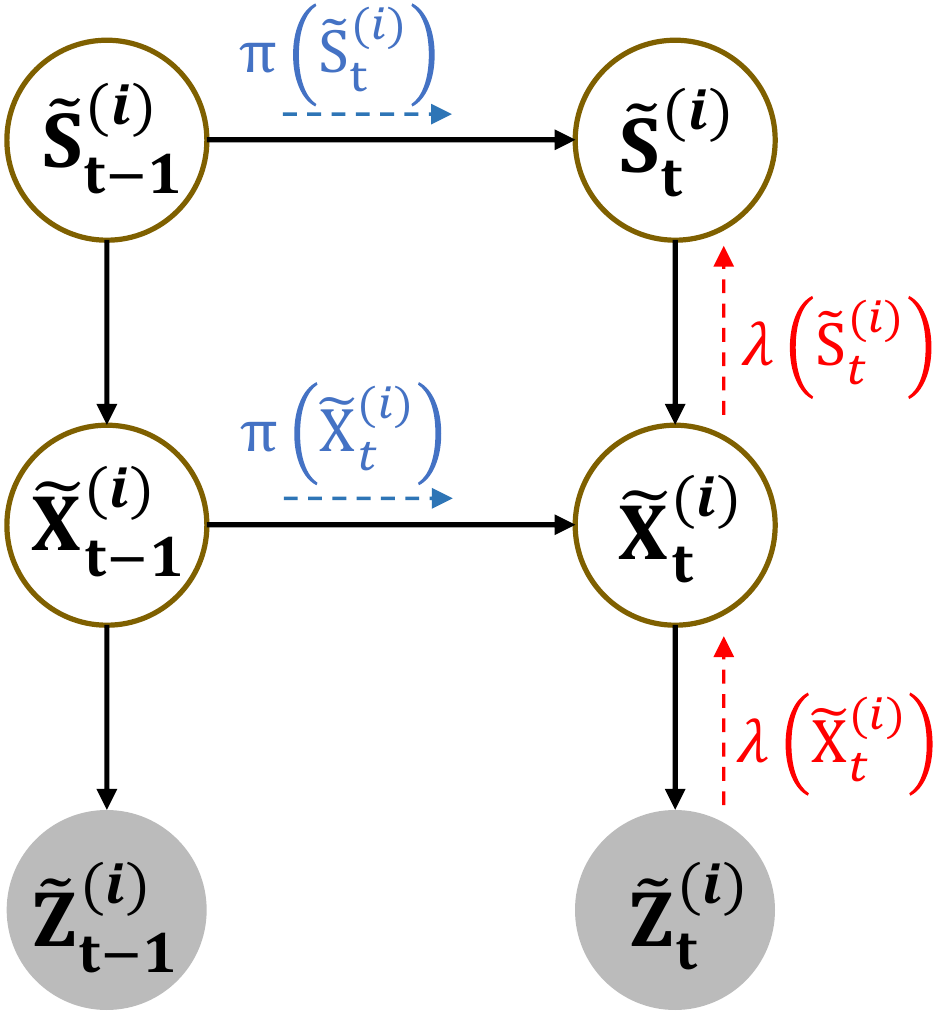}
        \\[-1.0mm]
        {\scriptsize (a)}
        \end{minipage}
        \begin{minipage}[b]{.50\linewidth}
            \centering
            \includegraphics[width=5.0cm]{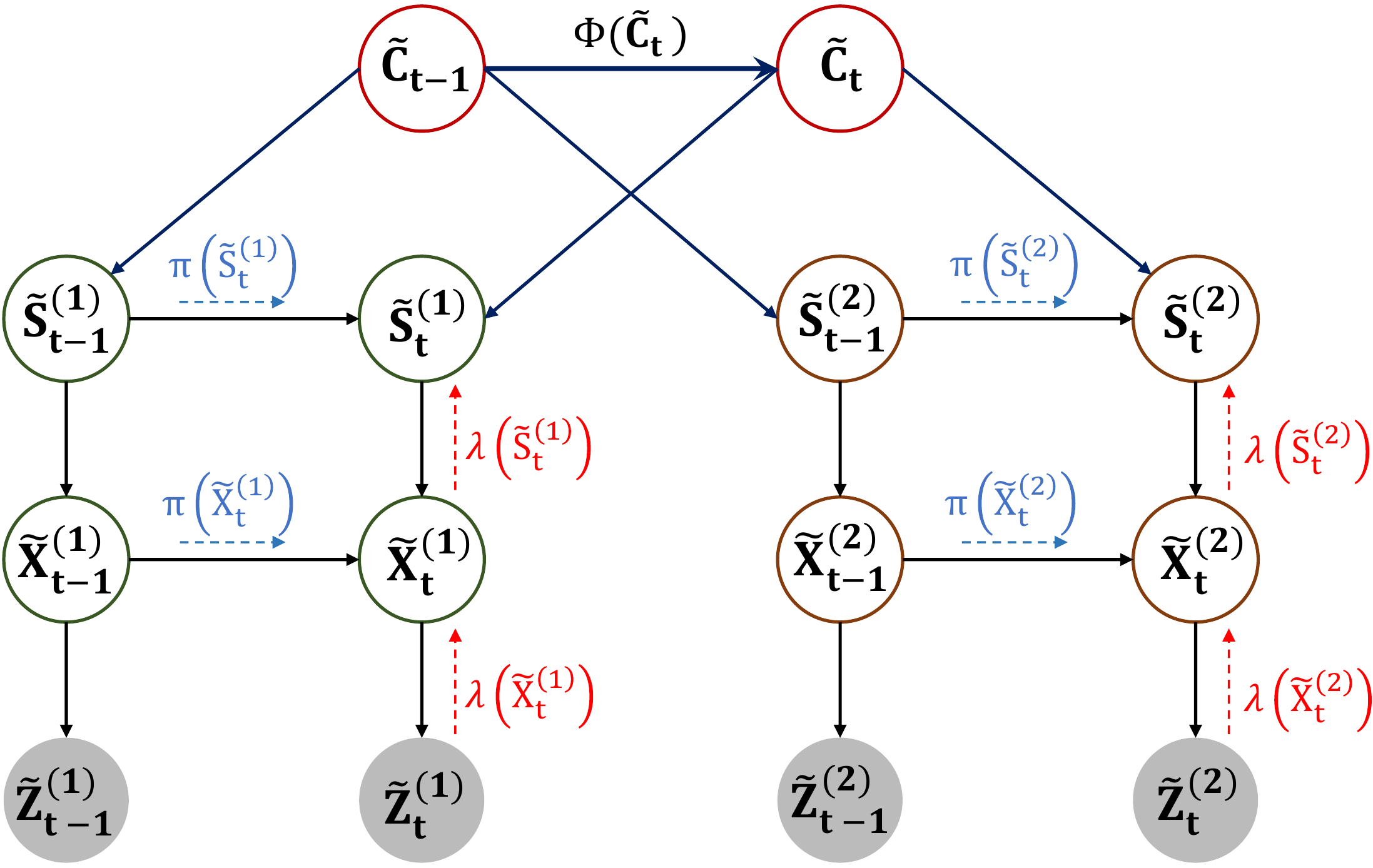}
            {\scriptsize (b)}
        \end{minipage}
        \caption{(a) The GDBN. (b) The coupled GDBN (C-GDBN) composed of two GDBNs representing the two signals received at the RSU where their discrete hidden variables are stochastically coupled.}
        \label{fig_GDBN_CGDBN}
    \end{center}
\end{figure}

The hierarchical dynamic models defined in \eqref{eq_discreteLevel}, \eqref{eq_continuousLevel} and \eqref{eq_observationLevel} are structured in a Generalized Dynamic Bayesian Network (GDBN) \cite{9858012} as shown in Fig.~\ref{fig_GDBN_CGDBN}-(a) that provides a probabilistic graphical model expressing the conditional dependencies among random hidden variables and observable states. The generative process explaining how sensory signals have been generated can be factorized as:
%
\begin{equation} \label{eq_generative_process}
\begin{split}
    \mathrm{P}(\mathrm{\tilde{Z}}_{t}^{(i)}, \mathrm{\tilde{X}}_{t}^{(i)}, \mathrm{\tilde{S}}_{t}^{(i)}) = \mathrm{P}(\mathrm{\tilde{S}}_{0}^{(i)}) \mathrm{P}(\mathrm{\tilde{X}}_{0}^{(i)}) \\ \bigg[ \prod_{t=1}^{\mathrm{T}} \mathrm{P}(\mathrm{\tilde{Z}}_{t}^{(i)}|\mathrm{\tilde{X}}_{t}^{(i)}) \mathrm{P}(\mathrm{\tilde{X}}_{t}^{(i)}|\mathrm{\tilde{X}}_{t-1}^{(i)}, \mathrm{\tilde{S}}_{t}^{(i)}) \mathrm{P}(\mathrm{\tilde{S}}_{t}^{(i)}|\mathrm{\tilde{S}}_{t-1}^{(i)}) \bigg],
\end{split}
\end{equation}
where $\mathrm{P}(\mathrm{\tilde{S}}_{0}^{(i)})$ and $\mathrm{P}(\mathrm{\tilde{X}}_{0}^{(i)})$ are initial prior distributions, $\mathrm{P}(\mathrm{\tilde{Z}}_{t}^{(i)}|\mathrm{\tilde{X}}_{t}^{(i)})$ is the likelihood, $\mathrm{P}(\mathrm{\tilde{X}}_{t}^{(i)}|\mathrm{\tilde{X}}_{t-1}^{(i)}, \mathrm{\tilde{S}}_{t}^{(i)})$ and $\mathrm{P}(\mathrm{\tilde{S}}_{t}^{(i)}|\mathrm{\tilde{S}}_{t-1}^{(i)})$ are the transition densities describing the temporal and hierarchical dynamics of the generalized state-space model.
%
%
The generative process defined in \eqref{eq_generative_process} indicates the cause-effect relationships the model impose on the random variables $\mathrm{\tilde{S}}_{t}^{(i)}$, $\mathrm{\tilde{X}}_{t}^{(i)}$ and $\mathrm{\tilde{Z}}_{t}^{(i)}$ forming a chain of causality describing how one state contributes to the production of another state which is represented by the link $\mathrm{\tilde{S}}_{t}^{(i)} \rightarrow \mathrm{\tilde{X}}_{t}^{(i)} \rightarrow \mathrm{\tilde{Z}}_{t}^{(i)}$.

The RSU starts perceiving the environment using a static assumption about the environmental states evolution by assuming that sensory signals are only subject to random noise. Hence, RSU predicts the RF signal (or vehciles trajectory) using the following simplified model:
%
%
$\mathrm{\tilde{X}}_{t}^{(i)} = \mathrm{A} \mathrm{\tilde{X}}_{t-1}^{(i)} + \mathrm{\tilde{w}}_{t}$, 
that differs from \eqref{eq_continuousLevel} in the control vector $\mathrm{U}_{\mathrm{\Tilde{S}_{t}}^{(i)}}$ which is supposed to be null, i.e., $\mathrm{U}_{\mathrm{\Tilde{S}_{t}}^{(i)}} = 0$ as the dynamic rules explaining how the environmental states evolve with time are not discovered yet.
Those rules can be discovered by exploiting the generalized errors (GEs), i.e., the difference between predictions and observations. The GEs projected into the measurement space are calculated as:
%
%
$\tilde{\varepsilon}_{\mathrm{\tilde{Z}}_{t}^{(i)}}^{} = \mathrm{\tilde{Z}}_{t}^{(i)} - \mathrm{H} \mathrm{\tilde{X}}_{t}^{(i)}$.
Projecting $\tilde{\varepsilon}_{\mathrm{\tilde{Z}}_t}^{}$ back into the generalized state space can be done as follows:
\begin{equation}\label{GE_continuousLevel_initialModel}
    \tilde{\varepsilon}_{\mathrm{\tilde{X}}_t}^{(i)} = \mathrm{H}^{-1}\tilde{\varepsilon}_{\mathrm{\tilde{Z}}_{t}^{(i)}}^{}=\mathrm{H}^{-1}(\mathrm{\tilde{Z}}_{t}^{(i)}-\mathrm{H}\mathrm{\tilde{X}}_{t}^{(i)}) = \mathrm{H}^{-1}\mathrm{\tilde{Z}}_{t}^{(i)} - \mathrm{\tilde{X}}_{t}^{(i)}.
\end{equation}
The GEs defined in \eqref{GE_continuousLevel_initialModel} can be grouped into discrete clusters in an unsupervised manner by employing the Growing Neural Gas (GNG). The latter produces a set of discrete variables (clusters) denoted by:
%
%
$\mathbf{\tilde{S}^{(i)}}=\{\mathrm{\tilde{S}}_{1}^{(i)},\mathrm{\tilde{S}}_{2}^{(i)},\dots,\mathrm{\tilde{S}}_{M_{i}}^{(i)}\}$,
where $M_{i}$ is the total number of clusters and each cluster $\mathrm{\tilde{S}}_{m}^{(i)} \in \mathbf{\tilde{S}^{(i)}}$ follows a Gaussian distribution composed of GEs with homogeneous properties, such that $\mathrm{\tilde{S}}_{m}^{(i)} \sim \mathcal{N}(\tilde{\mu}_{\mathrm{\tilde{S}}_{m}^{(i)}}=[\mu_{\tilde{S}_{m}^{(i)}}, \Dot{\mu}_{\tilde{S}_{m}^{(i)}}], \Sigma_{\mathrm{\tilde{S}}_{m}^{(i)}})$.
%
%
%
%
The dynamic transitions of the sensory signals among the available clusters can be captured in a time-varying transition matrix ($\Pi_{\tau}$) by estimating the time-varying transition probabilities $\pi_{ij}=\mathrm{P}(\mathrm{\tilde{S}}_{t}^{(i)}=i|\mathrm{\tilde{S}}_{t-1}^{(i)}=j, \tau)$ where $\tau$ is the time spent in $\mathrm{\tilde{S}}_{t-1}^{(i)}=j$ before transition to $\mathrm{\tilde{S}}_{t}^{(i)}=i$.

\subsection{Learning Coupled GDBN (C-GDBN)}
The learning procedure described in the previous section can be executed for each signal type, i.e., RF and GPS. After learning a separated GDBN model for each signal type, we analyse the interaction behaviour between RF signal and GPS signal received at the RSU by tracking the cluster firing among $\mathbf{\tilde{S}^{(1)}}$ and $\mathbf{\tilde{S}^{(2)}}$ during a certain experience. Such an interaction can be encoded in a Coupled GDBN (C-GDBN) as shown in Fig.\ref{fig_GDBN_CGDBN}-(b) composed of the two GDBNs representing the two signals where their hidden variables at the discrete level are stochastically coupled (in $\mathrm{\tilde{C}}_{t}{=}[\mathrm{\tilde{S}}_{t}^{(1)},\mathrm{\tilde{S}}_{t}^{(2)}]$) as those variables are uncorrelated but have coupled means.
The interactive matrix $\Phi \in \mathbb{R}^{M_{1},M_{2}}$ encodes the firing cluster pattern allowing to predict the GPS signal from RF signal is defined as follows:
\begin{equation} \label{interactiveTM_fromRFtoGPS}
\Phi = 
        \begin{bmatrix} 
            \mathrm{P}(\mathrm{\Tilde{S}_{1}}^{(2)}|\mathrm{\Tilde{S}_{1}}^{(1)}) & \mathrm{P}(\mathrm{\Tilde{S}_{2}}^{(2)}|\mathrm{\Tilde{S}_{1}}^{(1)}) & \dots & \mathrm{P}(\mathrm{\Tilde{S}_{M_{2}}}^{(2)}|\mathrm{\Tilde{S}_{1}}^{(1)}) \\
            \mathrm{P}(\mathrm{\Tilde{S}_{1}}^{(2)}|\mathrm{\Tilde{S}_{2}}^{(1)}) & \mathrm{P}(\mathrm{\Tilde{S}_{2}}^{(2)}|\mathrm{\Tilde{S}_{2}}^{(1)}) & \dots & \mathrm{P}(\mathrm{\Tilde{S}_{M_{2}}}^{(2)}|\mathrm{\Tilde{S}_{2}}^{(1)}) \\
            \vdots & \vdots & \ddots & \vdots \\
            \mathrm{P}(\mathrm{\Tilde{S}_{1}}^{(2)}|\mathrm{\Tilde{S}_{M_{1}}}^{(1)}) & \mathrm{P}(\mathrm{\Tilde{S}_{2}}^{(2)}|\mathrm{\Tilde{S}_{M_{1}}}^{(1)}) & \dots & \mathrm{P}(\mathrm{\Tilde{S}_{M_{2}}}^{(2)}|\mathrm{\Tilde{S}_{M_{1}}}^{(1)}) 
        \end{bmatrix}.
\end{equation}
%
%
%

\subsection{Joint Prediction and Perception}
RSU starts predicting the RF signals it expects to receive from each vehicle based on a Modified Markov Jump Particle Filter (M-MJPF) \cite{9858012} that combines Particle filter (PF) and Kalman filter (KF) to perform temporal and hierarchical predictions. Since the acquired C-GDBN allows predicting a certain signal's dynamic evolution based on another's evolution, it requires an interactive Bayesian filter capable of dealing with more complicated predictions. To this purpose, we propose to employ an Interactive M-MJPF (IM-MJPF) on the C-GDBN. The IM-MJPF consists of a PF that propagates a set of $L$ particles equally weighted, such that $\{\mathrm{\tilde{S}}_{t,l}^{(1)}, \mathrm{W}_{t,l}^{(1)}\}{\sim}\{\pi(\mathrm{\tilde{S}}_{t}^{(1)}), \frac{1}{L}\}$, where $\mathrm{\tilde{S}}_{t,l}^{(1)}$, $l \in L$ and $(.^{(1)})$ is the RF signal type. In addition, RSU relies on $\Phi$ defined in \eqref{interactiveTM_fromRFtoGPS} to predict $\mathrm{\tilde{S}}_{t}^{(2)}$ realizing the discrete cluster of vehicle's trajectory starting from the predicted RF signal according to: $\{\mathrm{\tilde{S}}_{t}^{(2)},\mathrm{W}_{t,l}^{(2)}\}{\sim} \{\Phi(\mathrm{\tilde{S}}_{t,l}^{(1)}){=}\mathrm{P}(.|\mathrm{\tilde{S}}_{t,l}^{(1)}), \mathrm{W}_{t,l}^{(2)}\}$. For each predicted discrete variable $\mathrm{\tilde{S}}_{t,l}^{(i)}$, a multiple KF is employed to predict multiple continuous variables which guided by the predictions at the higher level as declared in \eqref{eq_continuousLevel} that can be represented probabilistically as $\mathrm{P}(\mathrm{\tilde{X}}_{t}^{(i)}|\mathrm{\tilde{X}}_{t-1}^{(i)}, \mathrm{\tilde{S}}_{t}^{(i)})$. The posterior probability that is used to evaluate expectations is given by:
%
%
\begin{multline} \label{piX}
    \pi(\mathrm{\tilde{X}}_{t}^{(i)})=\mathrm{P}(\mathrm{\tilde{X}}_{t}^{(i)},\mathrm{\tilde{S}}_{t}^{(i)}|\mathrm{\tilde{Z}}_{t-1}^{(i)})= \\ \int \mathrm{P}(\mathrm{\tilde{X}}_{t}^{(i)}|\mathrm{\tilde{X}}_{t-1}^{(i)}, \mathrm{\tilde{S}}_{t}^{(i)}) \lambda(\mathrm{\tilde{X}}_{t-1}^{(i)})d\mathrm{\tilde{X}}_{t-1}^{(i)},
\end{multline}
where $\lambda(\mathrm{\tilde{X}}_{t-1}^{(i)}){=}\mathrm{P}(\mathrm{\tilde{Z}}_{t-1}^{(i)}|\mathrm{\tilde{X}}_{t-1}^{(i)})$. 
The posterior distribution can be updated (and so representing the updated belief) after having seen the new evidence $\mathrm{\tilde{Z}}_{t}^{(i)}$ by exploiting the diagnostic message $\lambda(\mathrm{\tilde{X}}_{t}^{(i)})$ in the following form: $\mathrm{P}(\mathrm{\tilde{X}}_{t}^{(i)}, \mathrm{\tilde{S}}_{t}^{(i)}|\mathrm{\tilde{Z}}_{t}^{(i)}) {=} \pi(\mathrm{\tilde{X}}_{t}^{(i)})\lambda(\mathrm{\tilde{X}}_{t}^{(i)})$. Likewise, belief in discrete hidden variables can be updated according to: $\mathrm{W}_{t,l}^{(i)}{=}\mathrm{W}_{t,l}^{(i)}\lambda (\mathrm{\tilde{S}}_{t}^{(i)})$ where:
%
%
$\lambda (\mathrm{\tilde{S}}_{t}^{(i)}) {=} \lambda (\mathrm{\Tilde{X}}_{t}^{(i)})\mathrm{P}(\mathrm{\Tilde{X}}_{t}^{(i)}|\mathrm{\tilde{S}}_{t}^{(i)}) {=} \mathrm{P}(\mathrm{\tilde{Z}}_{t}^{(i)}|\mathrm{\Tilde{X}}_{t}^{(i)})\mathrm{P}(\mathrm{\Tilde{X}}_{t}^{(i)}|\mathrm{\tilde{S}}_{t}^{(i)})$.

\subsection{Joint GPS spoofing and jamming detection}
RSU can evaluate the current situation and identify if V2I is under attack, or the satellite link is under spoofing based on a multiple abnormality indicator produced by the IM-MJPF. The first indicator calculates the similarity between the predicted RF signal and the observed one, which is defined as:
\begin{equation}\label{eq_CLA1}
    \Upsilon_{\mathrm{\tilde{X}}_{t}^{(1)}} = -ln \bigg( \mathcal{BC} \big(\pi(\mathrm{\tilde{X}}_{t}^{(1)}),\lambda(\mathrm{\tilde{X}}_{t}^{(1)}) \big) \bigg),
\end{equation}
where $\mathcal{BC}(.){=}\int \sqrt{\pi(\mathrm{\tilde{X}}_{t}^{(1)}),\lambda(\mathrm{\tilde{X}}_{t}^{(1)}})d\mathrm{\tilde{X}}_{t}^{(1)}$ is the Bhattacharyya coefficient.
The second indicator calculates the similarity between the predicted GPS signal (from the RF signal) and the observed one after decoding the RF signal which is defined as:
\begin{equation}\label{eq_CLA2}
    \Upsilon_{\mathrm{\tilde{X}}_{t}^{(2)}} = -ln \bigg( \mathcal{BC} \big(\pi(\mathrm{\tilde{X}}_{t}^{(2)}),\lambda(\mathrm{\tilde{X}}_{t}^{(2)}) \big) \bigg),
\end{equation}
where $\mathcal{BC}(.){=}\int \sqrt{\pi(\mathrm{\tilde{X}}_{t}^{(2)}),\lambda(\mathrm{\tilde{X}}_{t}^{(2)}})d\mathrm{\tilde{X}}_{t}^{(2)}$.
Different hypotheses can be identified by the RSU to understand the current situation whether there is: a jammer attacking the V2I link, or a spoofer attacking the link between the satellite and the vehicle or both jammer and spoofer are absent according to:
\begin{equation}
    \begin{cases}
        \mathcal{H}_{0}: \text{if} \ \ \Upsilon_{\mathrm{\tilde{X}}_{t}^{(1)}} < \xi_{1} \ \text{and} \ \Upsilon_{\mathrm{\tilde{X}}_{t}^{(2)}} < \xi_{2}, \\
        \mathcal{H}_{1}: \text{if} \ \ \Upsilon_{\mathrm{\tilde{X}}_{t}^{(1)}} \geq \xi_{1} \ \text{and} \ \Upsilon_{\mathrm{\tilde{X}}_{t}^{(2)}} \geq \xi_{2}, \\
        \mathcal{H}_{2}: \text{if} \ \ \Upsilon_{\mathrm{\tilde{X}}_{t}^{(1)}} < \xi_{1} \ \text{and} \ \Upsilon_{\mathrm{\tilde{X}}_{t}^{(2)}} \geq \xi_{2},
    \end{cases}
\end{equation}
%
%
%
where $\xi_{1} = \mathbb{E}[\Bar{\Upsilon}_{\mathrm{\tilde{X}}_{t}^{(1)}}] + 3\sqrt{\mathbb{V}[\Bar{\Upsilon}_{\mathrm{\tilde{X}}_{t}^{(1)}}]}$, and $\xi_{2} = \mathbb{E}[\Bar{\Upsilon}_{\mathrm{\tilde{X}}_{t}^{(2)}}] + 3\sqrt{\mathbb{V}[\Bar{\Upsilon}_{\mathrm{\tilde{X}}_{t}^{(2)}}]}$. In $\xi_{1}$ and $\xi_{2}$, $\Bar{\Upsilon}_{\mathrm{\tilde{X}}_{t}^{(1)}}$ and $\Bar{\Upsilon}_{\mathrm{\tilde{X}}_{t}^{(2)}}$ stand for the abnormality signals during training (i.e., normal situation when jammer and spoofer are absent).

\subsection{Evaluation metrics}
In order to evaluate the performance of the proposed method to jointly detect jammer and GPS spoofer, we adopt the jammer detection probability ($\mathrm{P}_{d}^{j}$) and the spoofer detection probability ($\mathrm{P}_{d}^{s}$), respectively, which are defined as:
\begin{equation}
    \mathrm{P}_{d}^{j} = \mathrm{Pr}(\Upsilon_{\mathrm{\tilde{X}}_{t}^{(1)}}\geq \xi_{1}, \Upsilon_{\mathrm{\tilde{X}}_{t}^{(2)}} \geq \xi_{2}|\mathcal{H}_{1}),
\end{equation}
%
%
\begin{equation}
    \mathrm{P}_{d}^{s} = \mathrm{Pr}(\Upsilon_{\mathrm{\tilde{X}}_{t}^{(1)}}< \xi_{1}, \Upsilon_{\mathrm{\tilde{X}}_{t}^{(2)}} \geq \xi_{2}|\mathcal{H}_{2}).
\end{equation}
Also, we evaluate the accuracy of the proposed method in predicting and estimating the vehicles' trajectories and the expected RF signals by adopting the root mean square error (RMSE) defined as:
\begin{equation}
    RMSE = \sqrt{ \frac{1}{T} \sum_{t=1}^{T}\bigg( \mathrm{\tilde{Z}}_{t}^{(i)}-\mathrm{\tilde{X}}_{t}^{(i)} \bigg)^{2} },
\end{equation}
where $T$ is the total number of predictions.

\section{Simulation Results}
%
%
In this section, we evaluate the performance of the proposed method to jointly detect the jammer and the spoofer using extensive simulations. We consider $\mathrm{N}=2$ vehicles interacting inside the environment and exchanging their states (i.e., position and velocity) with the RSU. The vehicles move along predefined trajectories performing various maneuvers which are picked from the \textit{Lankershim} dataset proposed by \cite{5206559}. The dataset depicts a four way intersection and includes about $19$ intersection maneuvers. RSU assigns one subchannel realizing the V2I link for each vehicle over which the vehicles' states are transmitted. The transmitted signal carrying the vehicle's state and the jamming signal are both QPSK modulated. 
The simulation settings are: carrier frequency of $2$GHz, BW${=}1.4$MHz, cell radius of $500$m, RSU antenna height and gain is $25$m and $8$ dBi, receiver noise figure of $5$dB, vehicle antenna height and gain is $1.5$m and $3$dBi, vehicle speed is $40$Km/h, V2I transmit power is $23$dBm, jammer transmit power ranging from $20$dBm to $40$dBm, SNR of $20$dB, path loss model ($128.1{+}37.6log d$), Log-normal shadowing with $8$dB standard deviation and a fast fading channel following the Rayleigh distribution.
\begin{figure}[ht!]
    \begin{center}
        \begin{minipage}[b]{.55\linewidth}
        \centering
            \includegraphics[width=5.0cm]{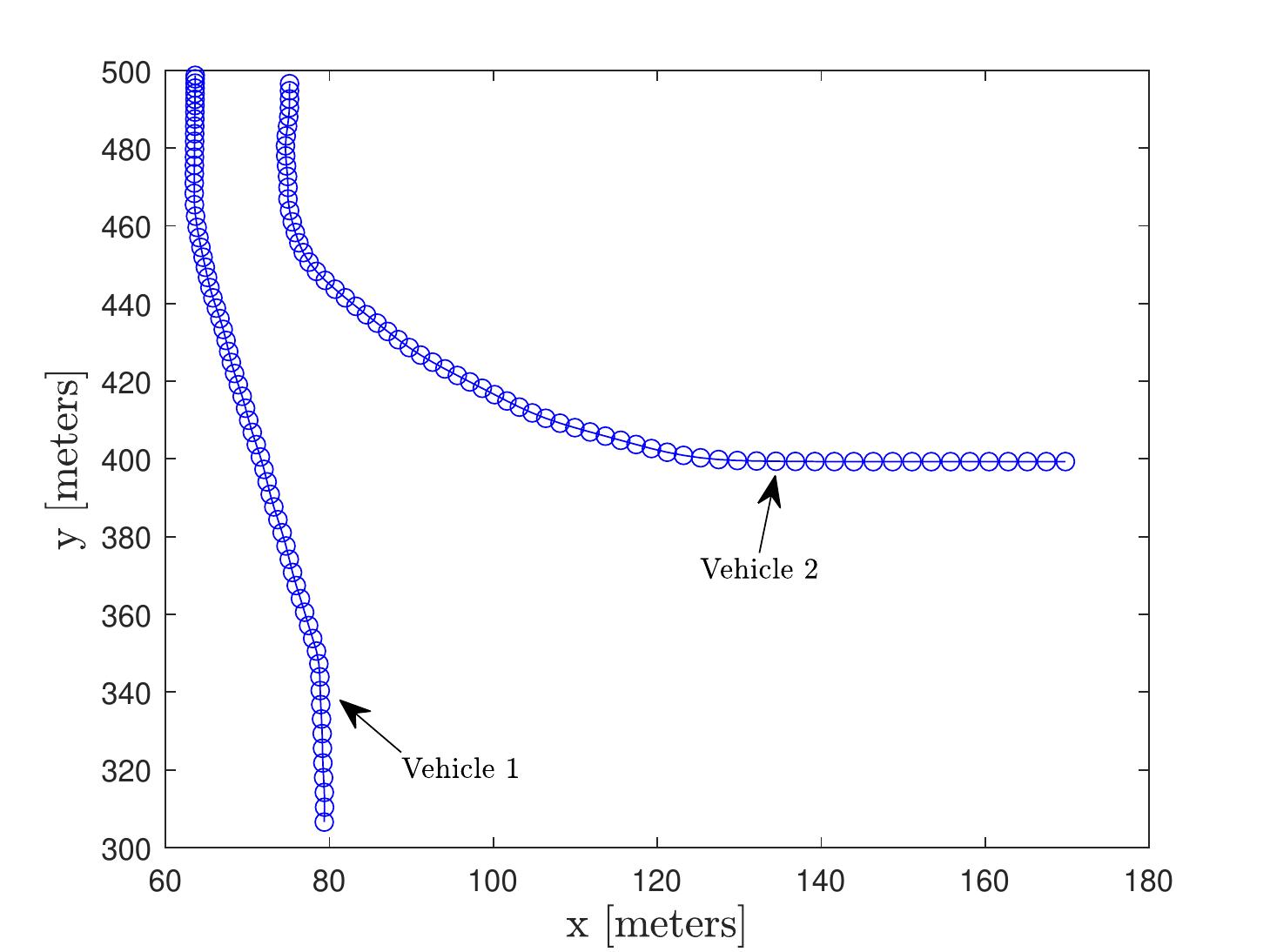}
        \\[-1.5mm]
        {\scriptsize (a)}
        \end{minipage}
        \begin{minipage}[b]{.49\linewidth}
            \centering
            \includegraphics[width=4.9cm]{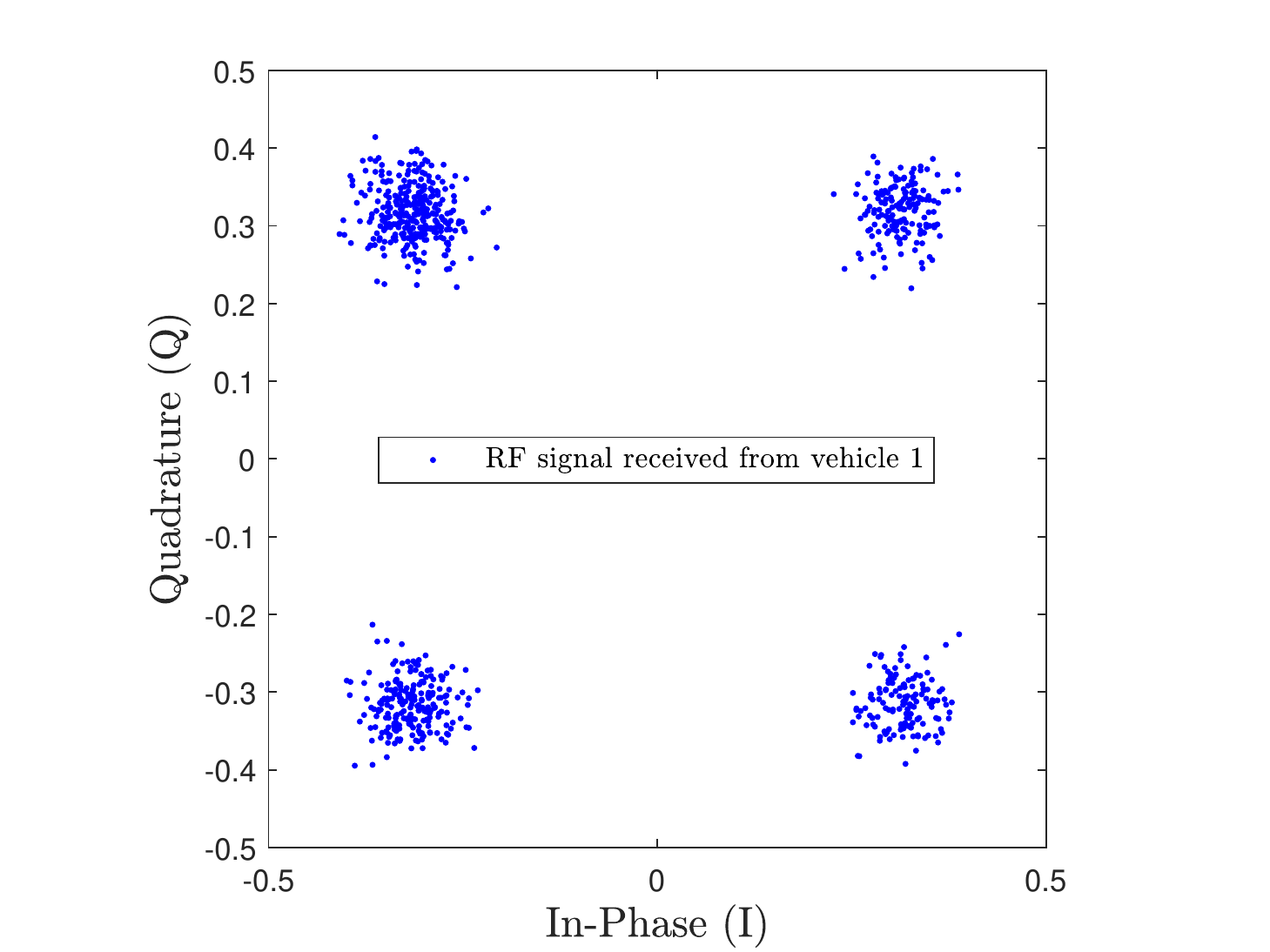}
            \\[-1.5mm]
            {\scriptsize (b)}
        \end{minipage}
        \begin{minipage}[b]{.49\linewidth}
            \centering
            \includegraphics[width=4.9cm]{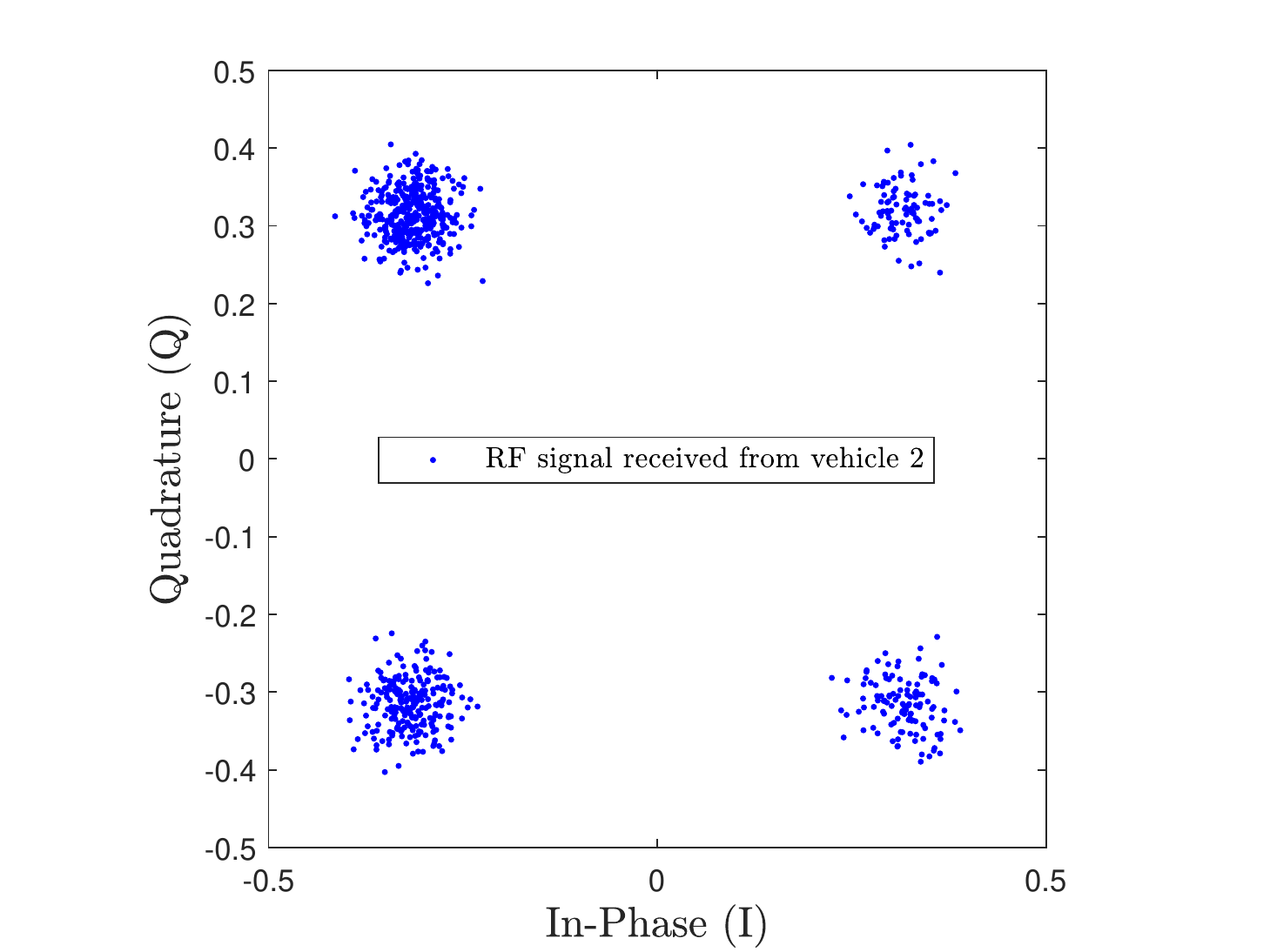}
            \\[-1.5mm]
            {\scriptsize (c)}
        \end{minipage}
        \caption{An example visualizing the received RF signals from the two vehicles and the corresponding trajectories: (a) Vehicles' trajectories, (b) received RF signal from vehicle 1, (c) received RF signal from vehicle 2.}
        \label{fig_receivedRFsignalandTrajectory}
    \end{center}
\end{figure}
\begin{figure}[t!]
    \begin{center}
        \begin{minipage}[b]{.49\linewidth}
            \centering
                \includegraphics[width=4.5cm]{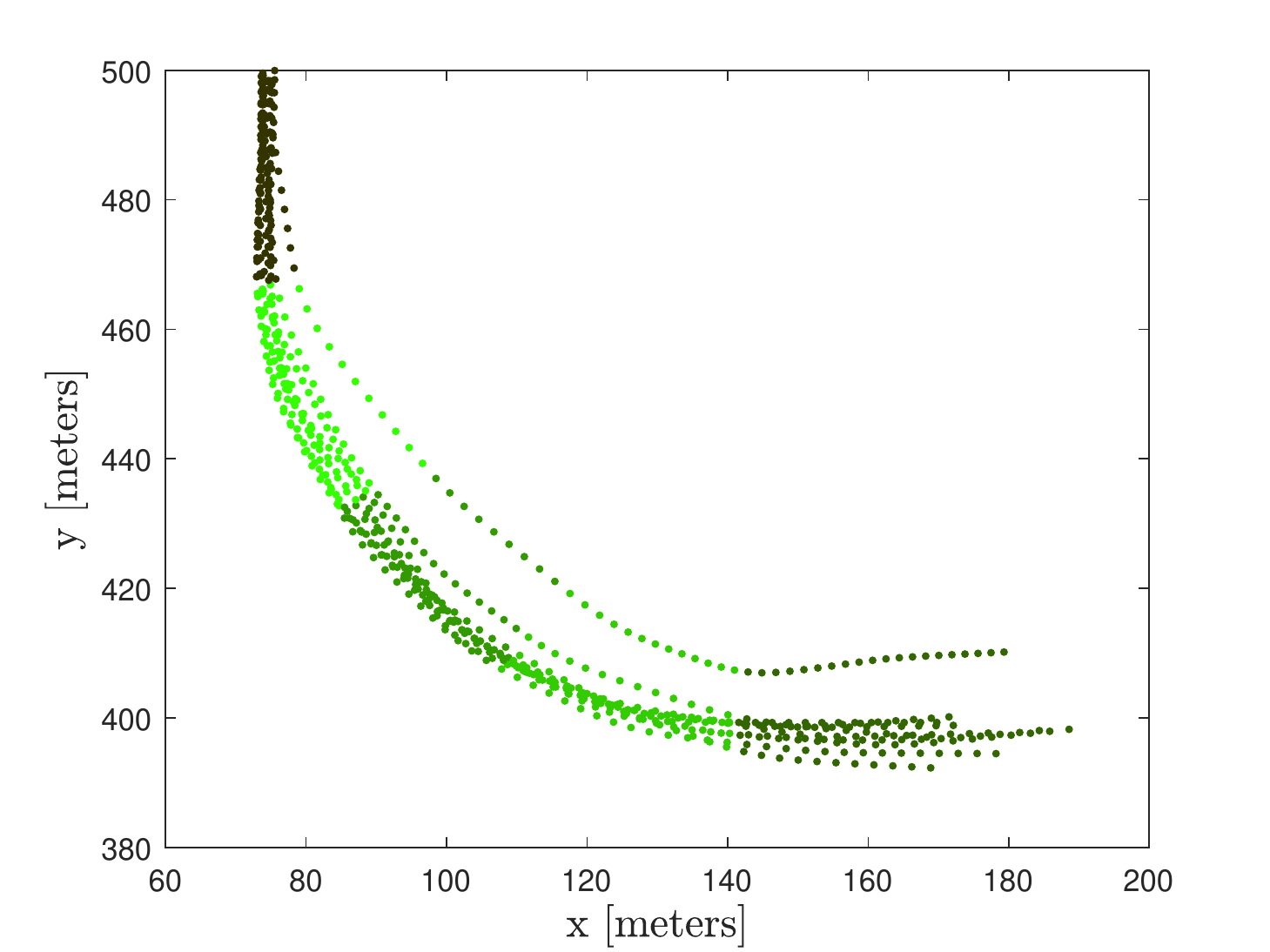}
            \\[-1.5mm]
            {\scriptsize (a)}
        \end{minipage}
        \begin{minipage}[b]{.49\linewidth}
            \centering
                \includegraphics[width=4.5cm]{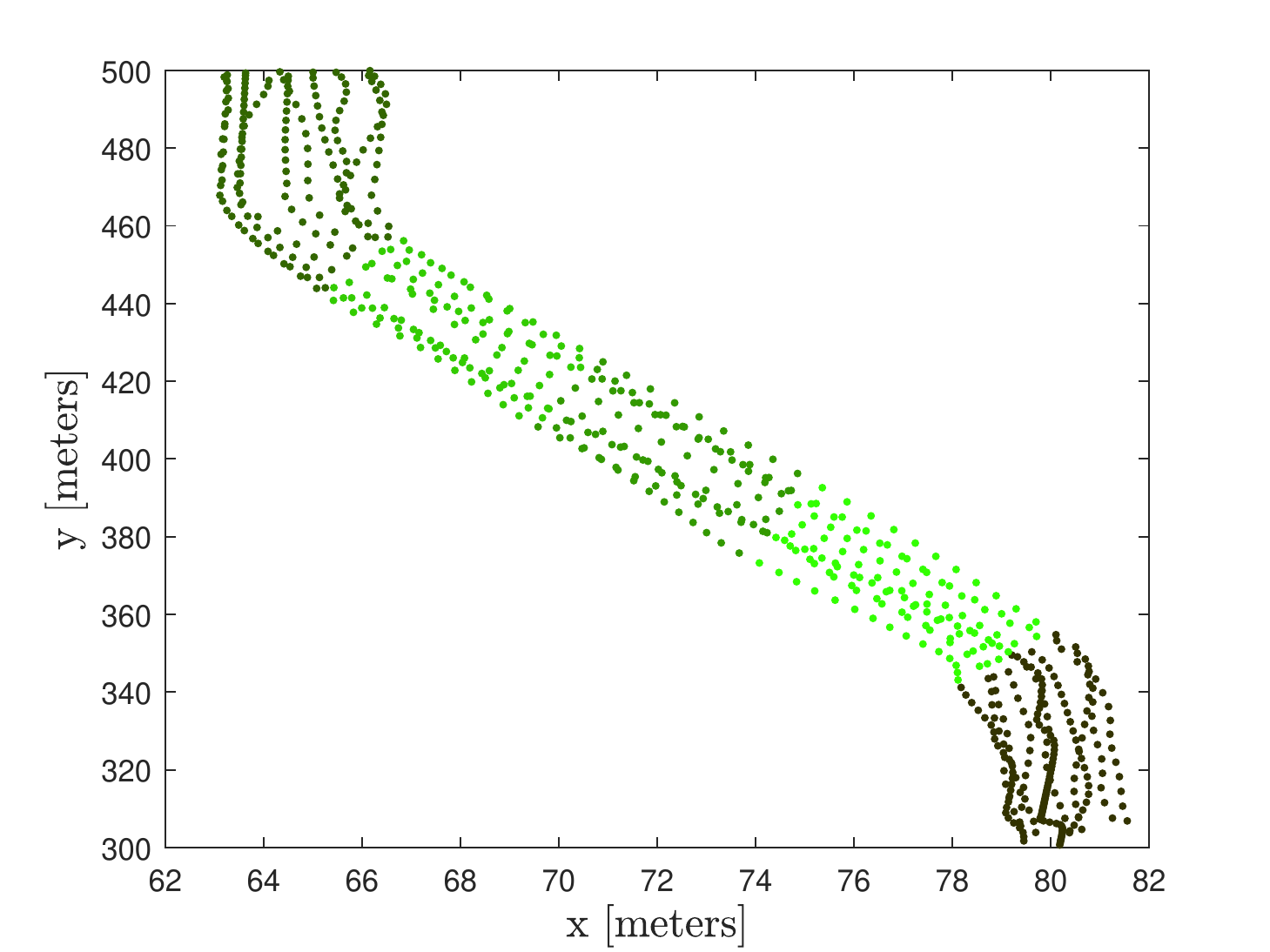}
            \\[-1.5mm]
            {\scriptsize (b)}
        \end{minipage}
        \begin{minipage}[b]{0.49\linewidth}
            \centering
            \includegraphics[width=4.5cm]{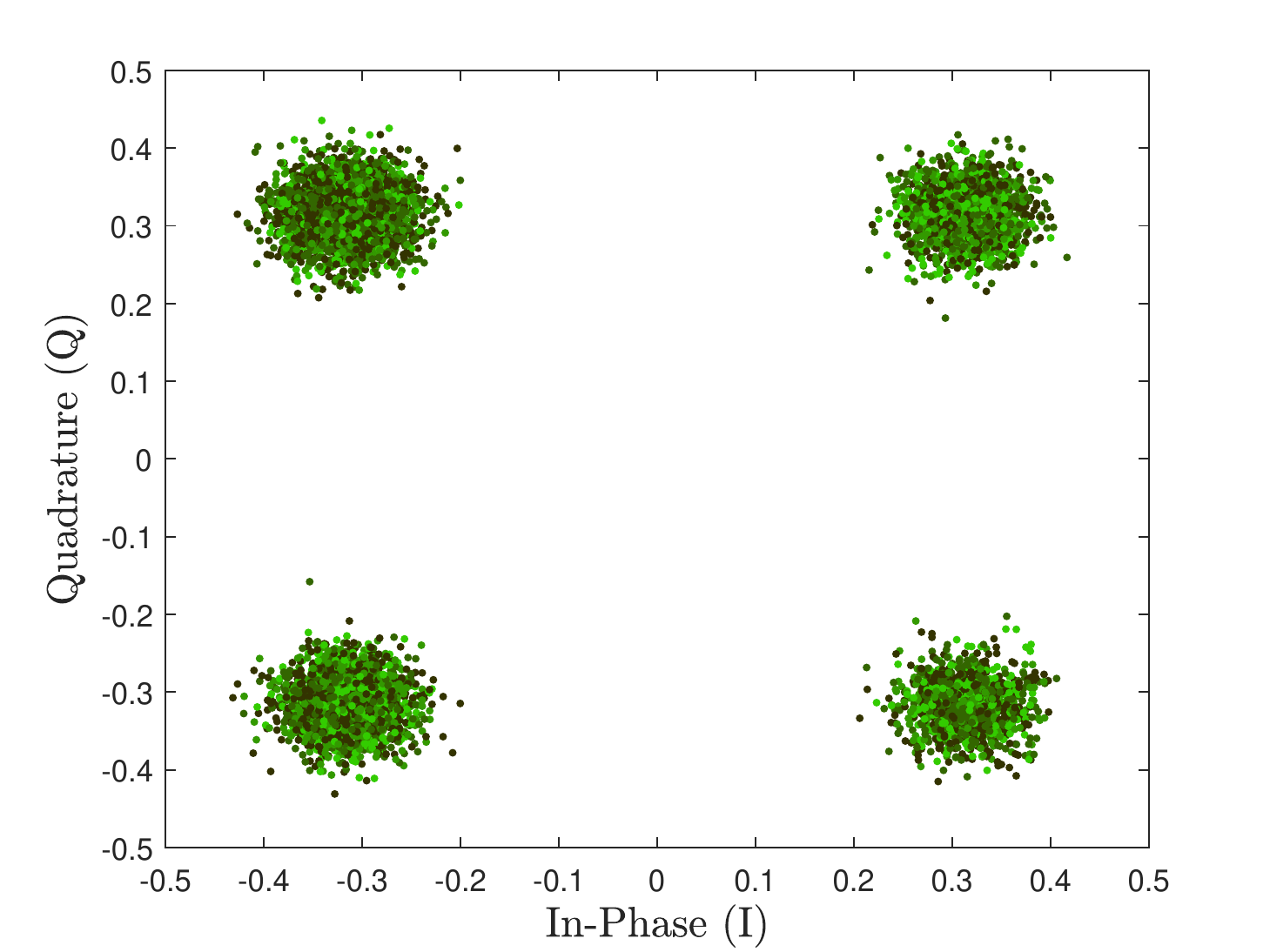}
            \\[-1.5mm]
            {\scriptsize (c)}
        \end{minipage}
        \begin{minipage}[b]{0.49\linewidth}
            \centering
            \includegraphics[width=4.5cm]{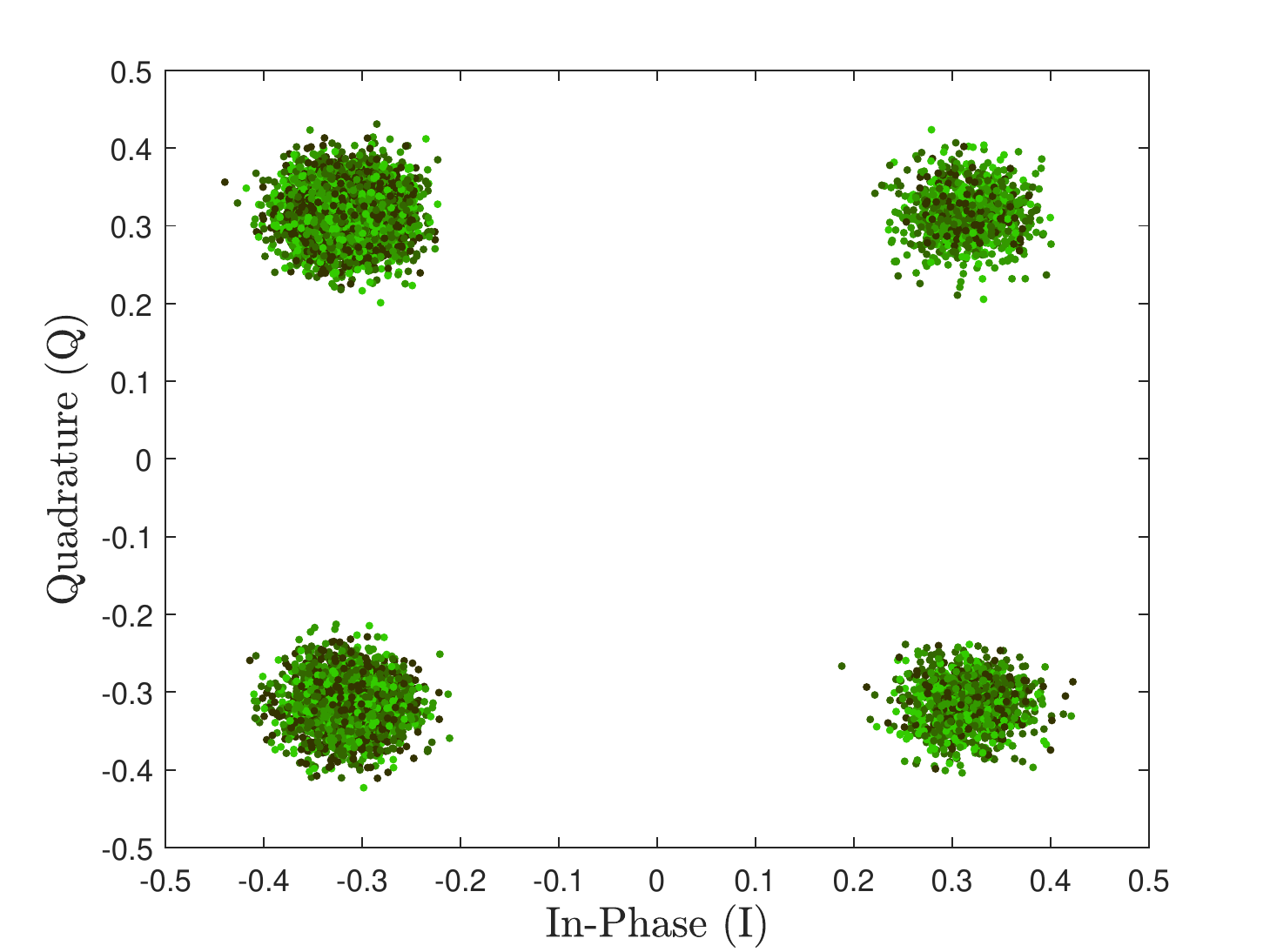}
            \\[-1.5mm]
            {\scriptsize (d)}
        \end{minipage}
        \caption{GNG output after clustering the generalized errors obtained from different experiences: (a) clustered trajectory of vehicle 1, (b) clustered trajectory of vehicle 2, (c) clustered RF signal received from vehicle 1, (d) clustered RF signal received from vehicle 2.}
        \label{fig_GNG_of_receivedRFsignalandTrajectory}
    \end{center}
\end{figure}

The RSU aims to learn multiple interactive models (i.e., C-GDBN models) encoding the cross relationship between the received RF signal from each vehicle and its corresponding trajectory. These models allow the RSU to predict the trajectory the vehicle will follow based on the received RF signal and evaluate whether the V2I is under jamming attacks or the satellite link is under spoofing. It is to note that the RSU is receiving only the RF signals from the two vehicles and obtaining their positions after decoding the RF signals. Thus, the RSU should be able to evaluate if the received RF signals are evolving according to the dynamic rules learned so far and if the vehicles are following the expected (right) trajectories to decide whether the V2I links are really under attack or whether the satellite link is under spoofing.

\begin{figure}[t!]
    \begin{center}
        \begin{minipage}[b]{.49\linewidth}
            \centering
                \includegraphics[width=4.5cm]{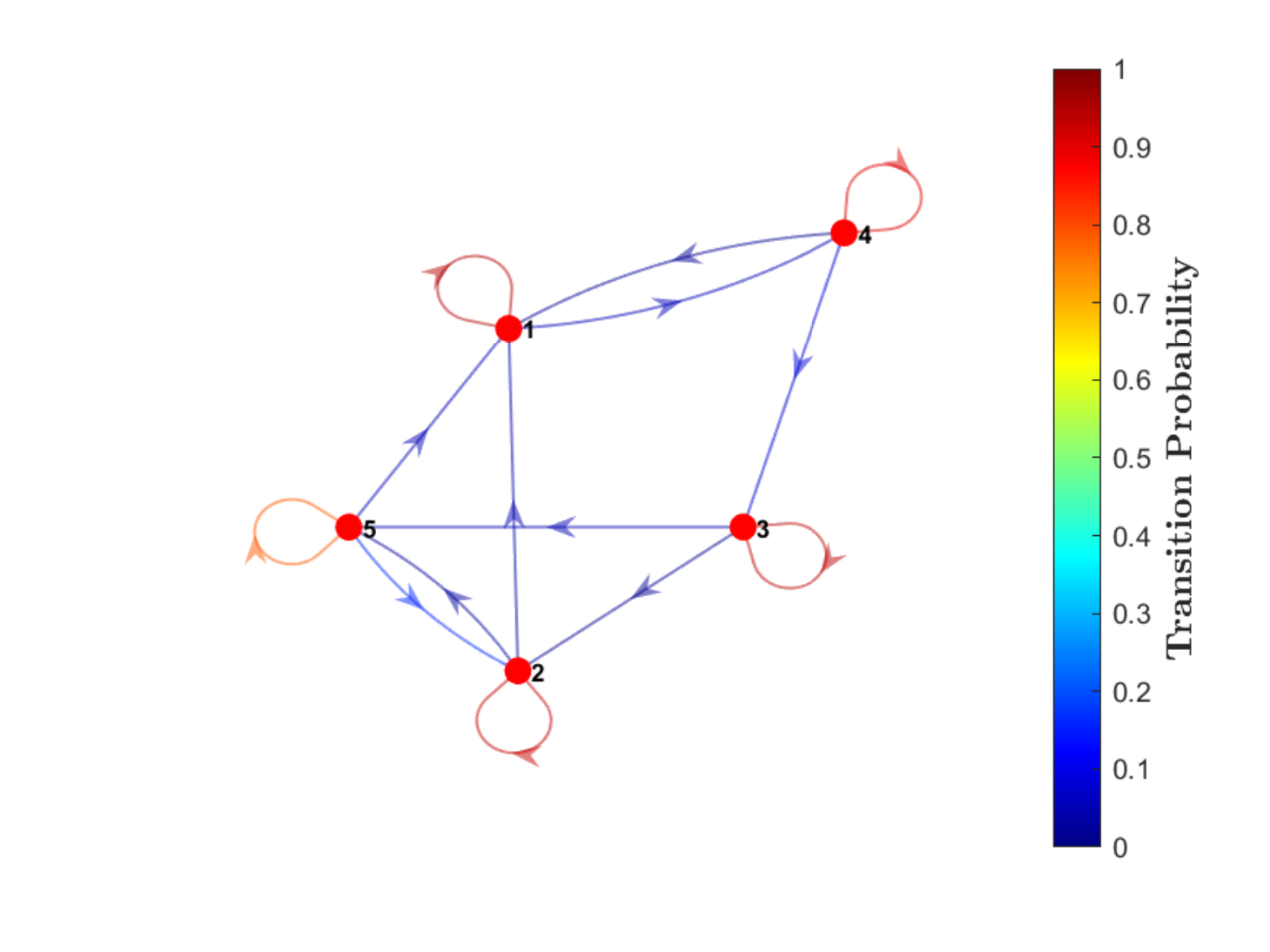}
            \\[-1.5mm]
            {\scriptsize (a)}
        \end{minipage}
        \begin{minipage}[b]{.49\linewidth}
            \centering
                \includegraphics[width=4.5cm]{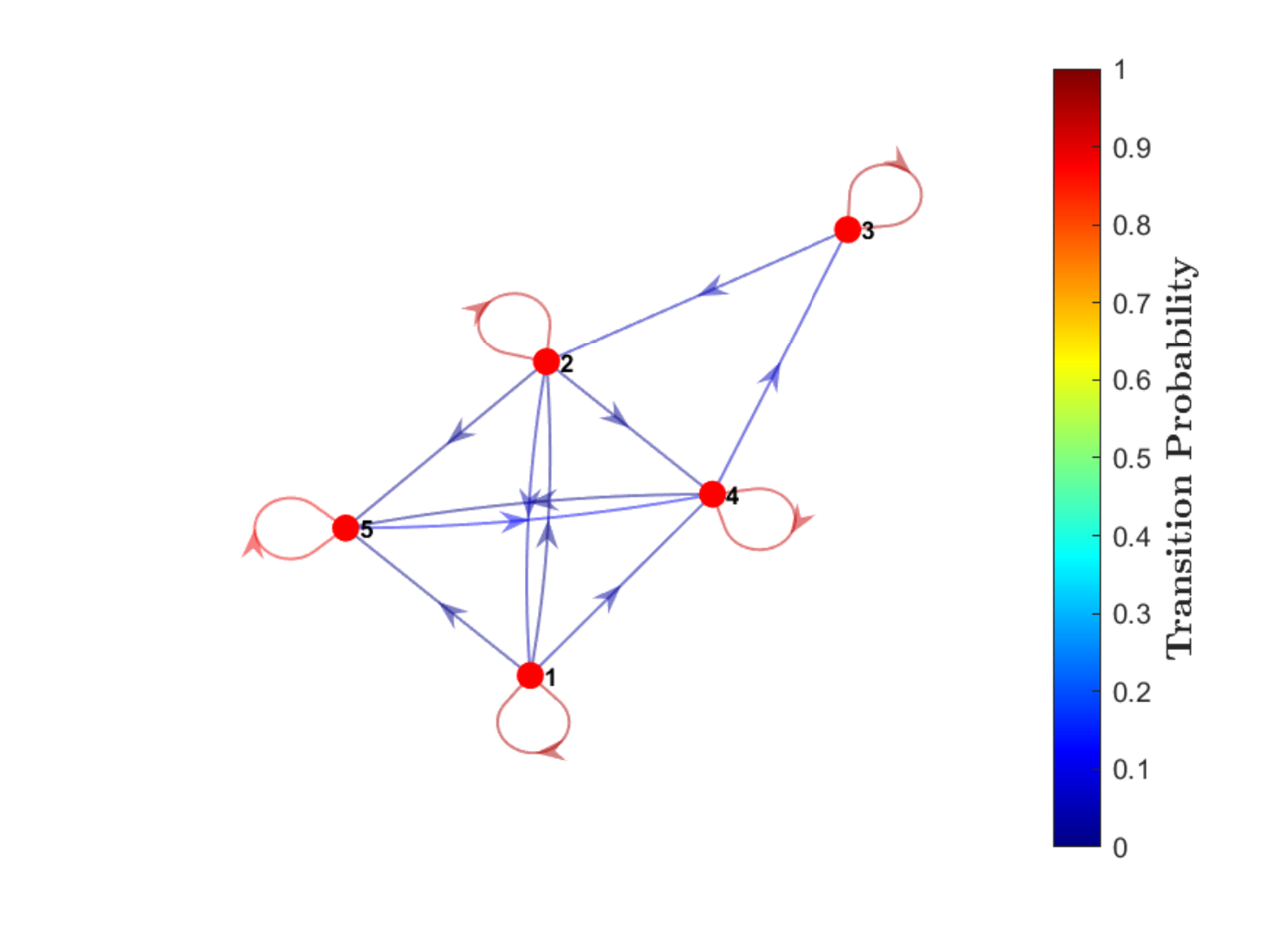}
            \\[-1.5mm]
            {\scriptsize (b)}
        \end{minipage}
        \begin{minipage}[b]{0.49\linewidth}
            \centering
            \includegraphics[width=4.5cm]{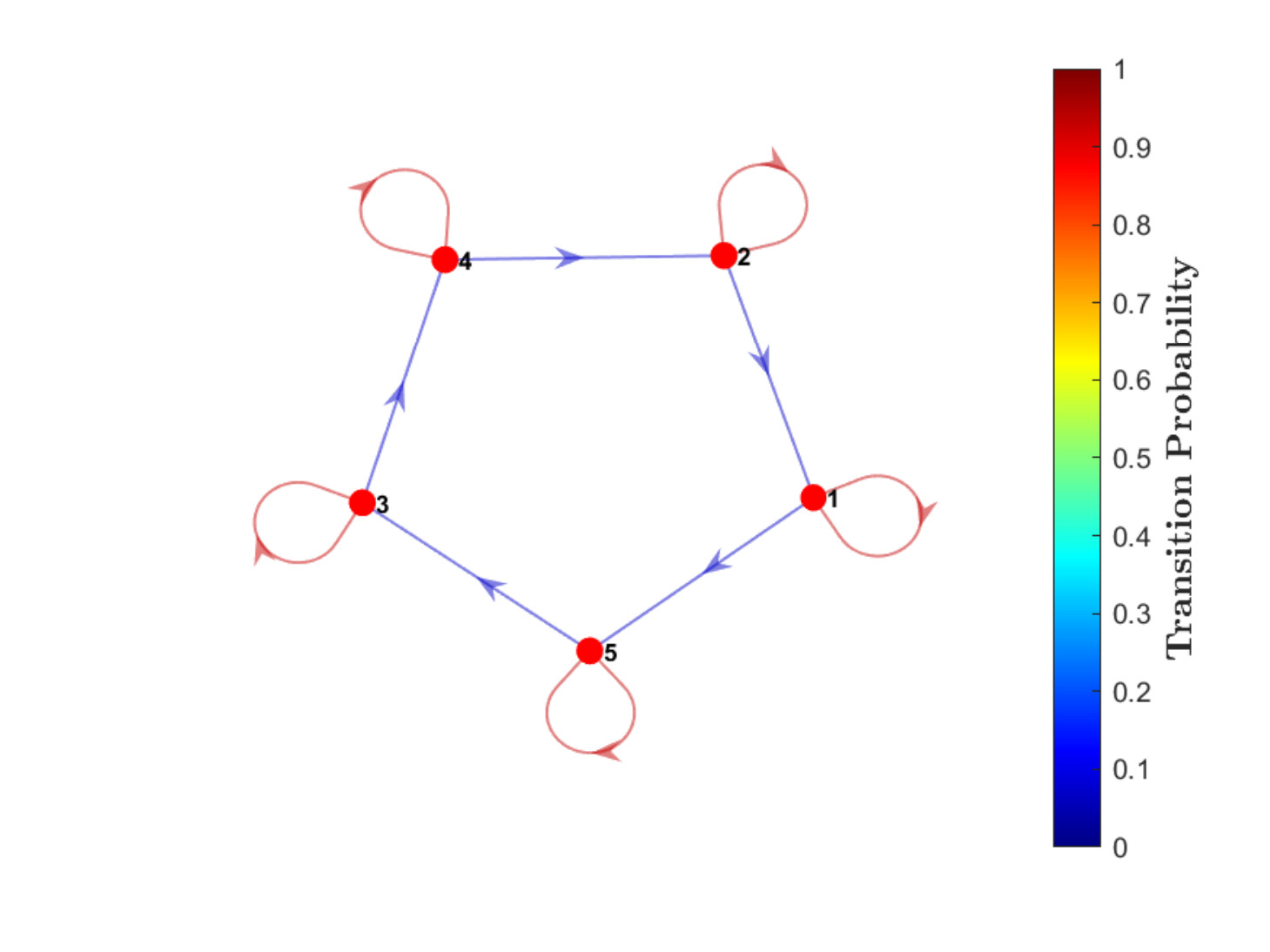}
            \\[-1.5mm]
            {\scriptsize (c)}
        \end{minipage}
        \begin{minipage}[b]{0.49\linewidth}
            \centering
            \includegraphics[width=4.5cm]{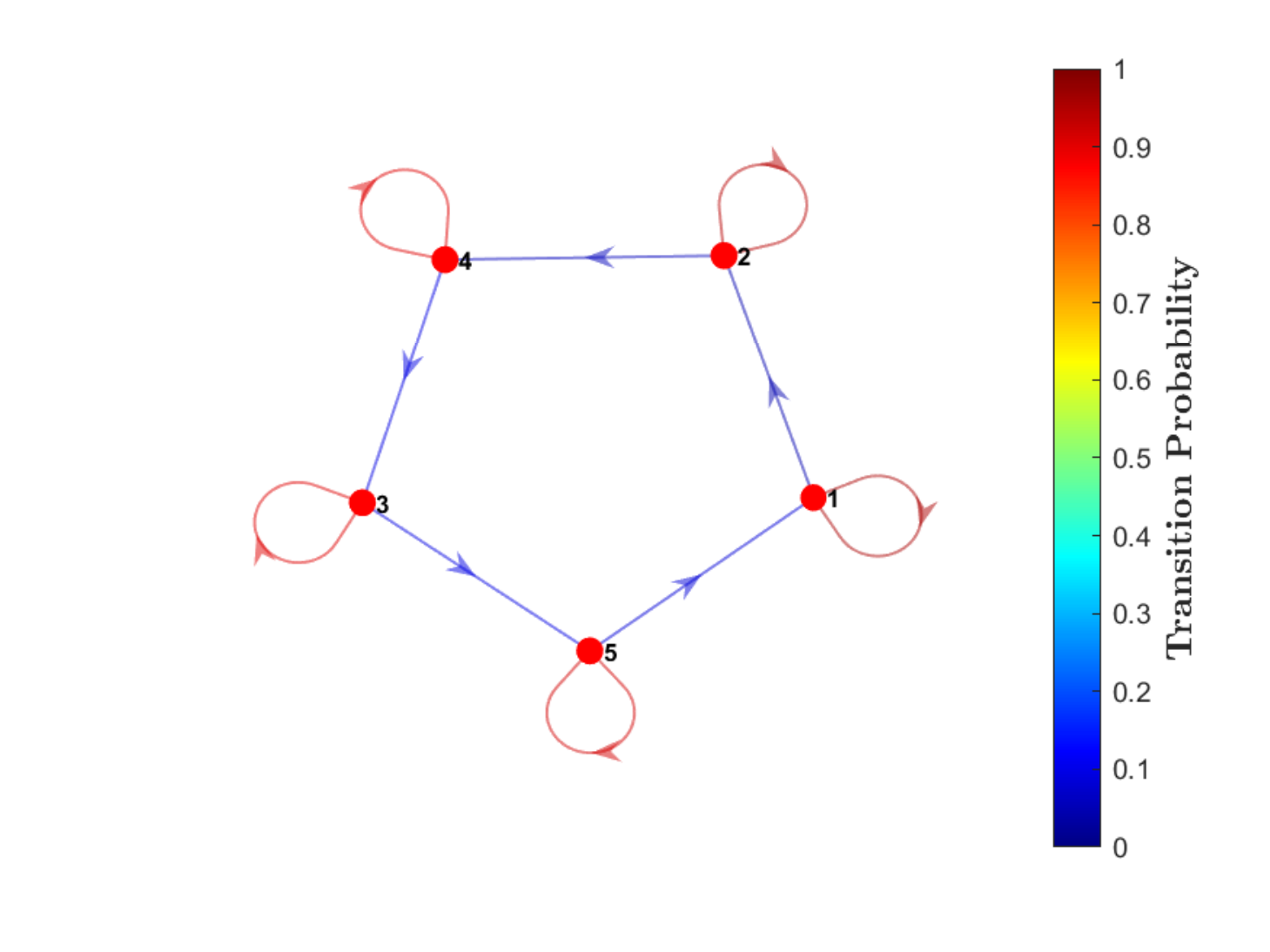}
            \\[-1.5mm]
            {\scriptsize (d)}
        \end{minipage}
        \caption{Graphical representation of the transition matrices (TM): (a) TM related to the trajectory of vehicle 1, (b) TM related to the trajectory of vehicle 2, (c) TM related to the RF signal received from vehicle 1, (d) TM related to the RF signal received from vehicle 2.}
        \label{fig_graphicalRep_transitionMatrices}
    \end{center}
\end{figure}

\begin{figure}[t!]
    \begin{center}
        \begin{minipage}[b]{.49\linewidth}
            \centering
                \includegraphics[width=3.8cm]{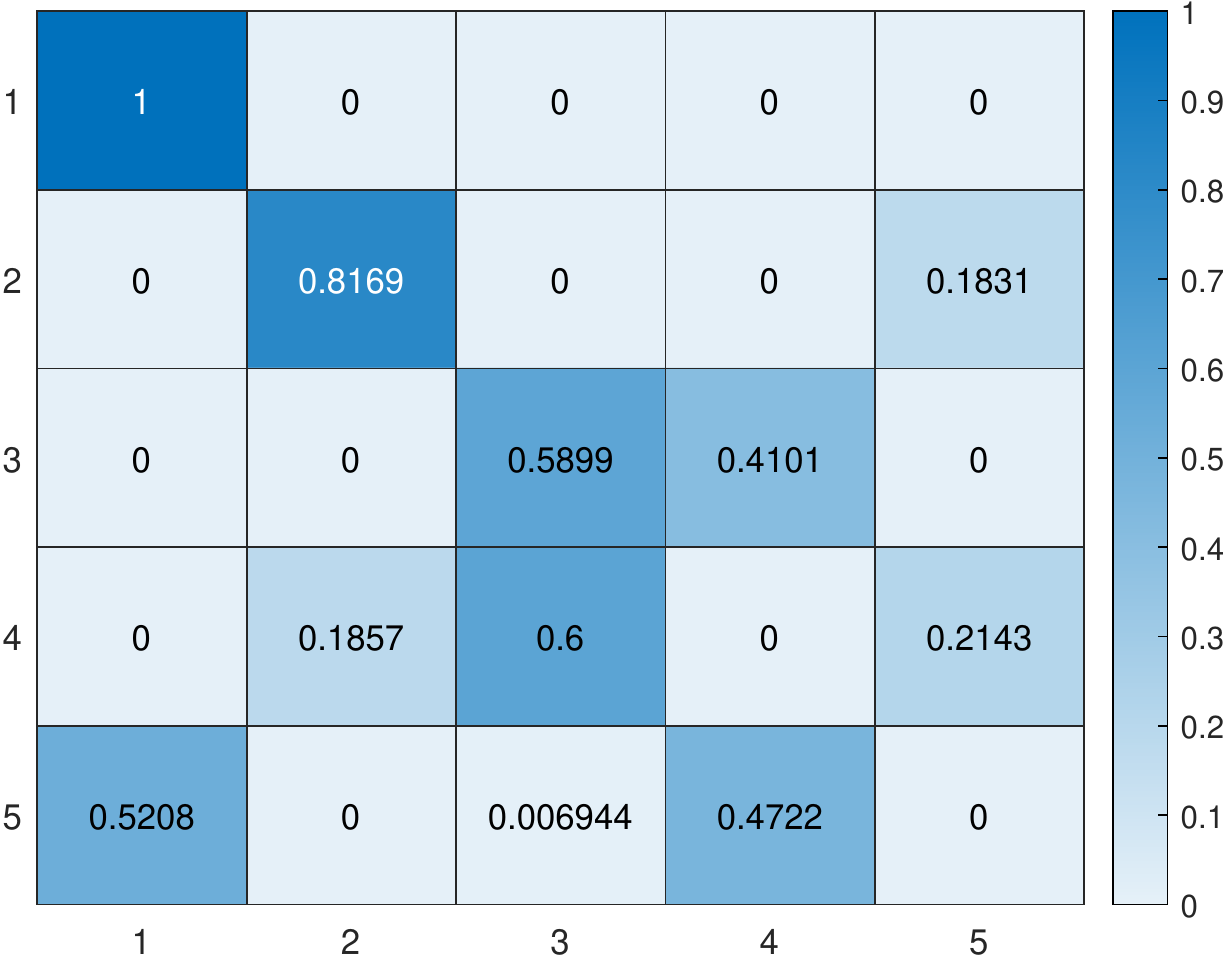}
            \\[-1.0mm]
            {\scriptsize (a)}
        \end{minipage}
        \begin{minipage}[b]{0.49\linewidth}
            \centering
            \includegraphics[width=3.8cm]{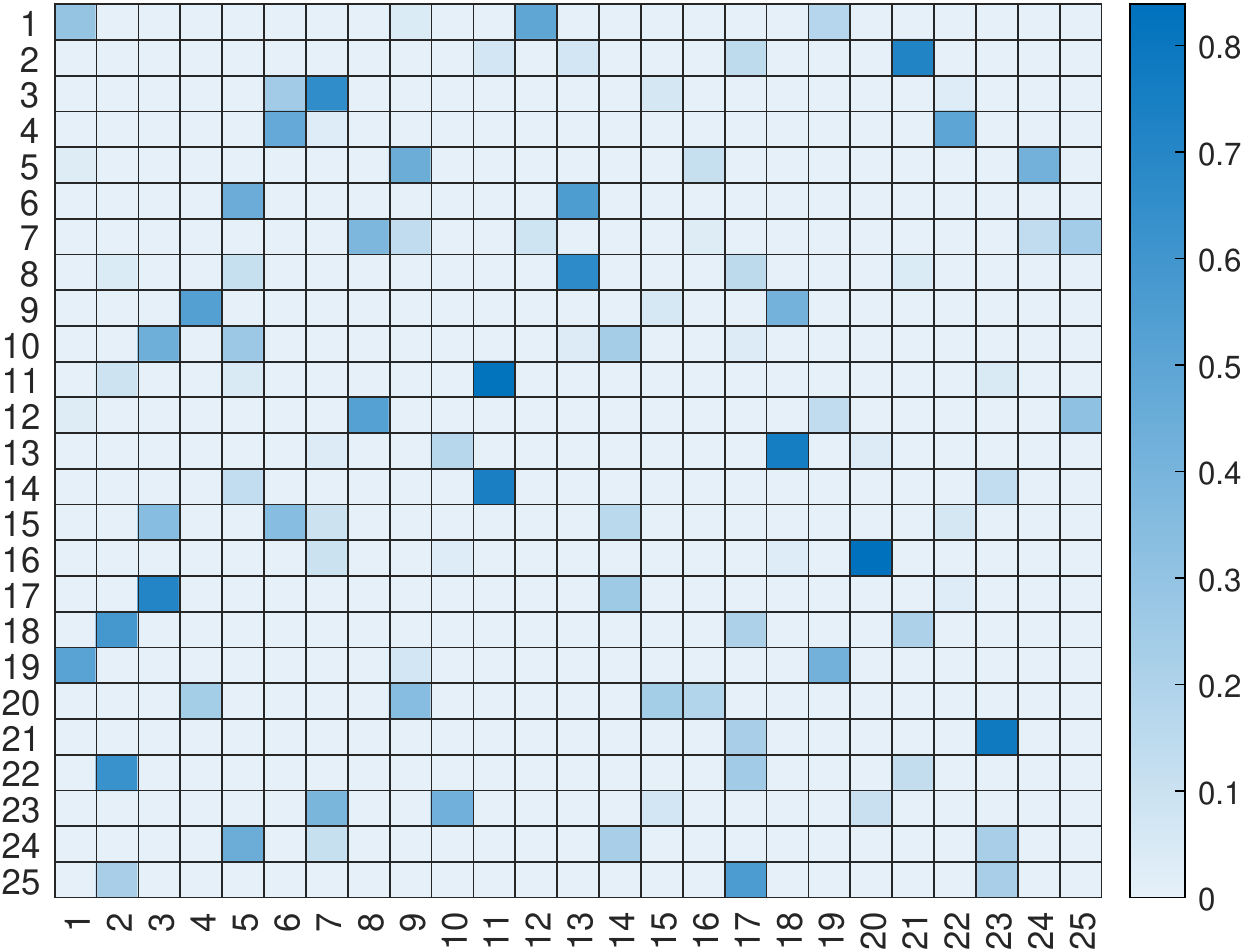}
            \\[-1.0mm]
            {\scriptsize (d)}
        \end{minipage}
        \caption{Interactive transition matrix defined in \eqref{interactiveTM_fromRFtoGPS} using different configurations: (a) $\mathrm{M_{1}}=5$, $\mathrm{M_{2}}=5$, (b) $\mathrm{M_{1}}=25$, $\mathrm{M_{2}}=25$.}
        \label{fig_interactiveMatrices}
    \end{center}
\end{figure}

Fig.~\ref{fig_receivedRFsignalandTrajectory}-(a) illustrates an example of the interaction between the two vehicles performing a particular manoeuvre, and Fig.~\ref{fig_receivedRFsignalandTrajectory}-(b) shows the received RF signals by the RSU from the two vehicles. At the beginning of the learning process, RSU performs predictions according to the simplified model defined in \eqref{eq_continuousLevel} where $\mathrm{U}_{\mathrm{\Tilde{S}_{t}}^{(i)}} {=} 0$.
After obtaining the generalized errors as pointed out in \eqref{GE_continuousLevel_initialModel}, RUS clusters those errors using GNG to learn two GDBN models encoding the dynamic rules of how the RF signal and the GPS signal evolve with time, respectively, as showed in Fig.~\ref{fig_GNG_of_receivedRFsignalandTrajectory} and Fig.~\ref{fig_graphicalRep_transitionMatrices}. RSU can couple the two GDBNs by learning the interactive transition matrix that is encoded in a C-GDBN as shown in Fig.~\ref{fig_interactiveMatrices}.
%
%
%
%
%
\begin{figure}[t!]
    \begin{center}
        \begin{minipage}[b]{.49\linewidth}
        \centering
            \includegraphics[width=4.9cm]{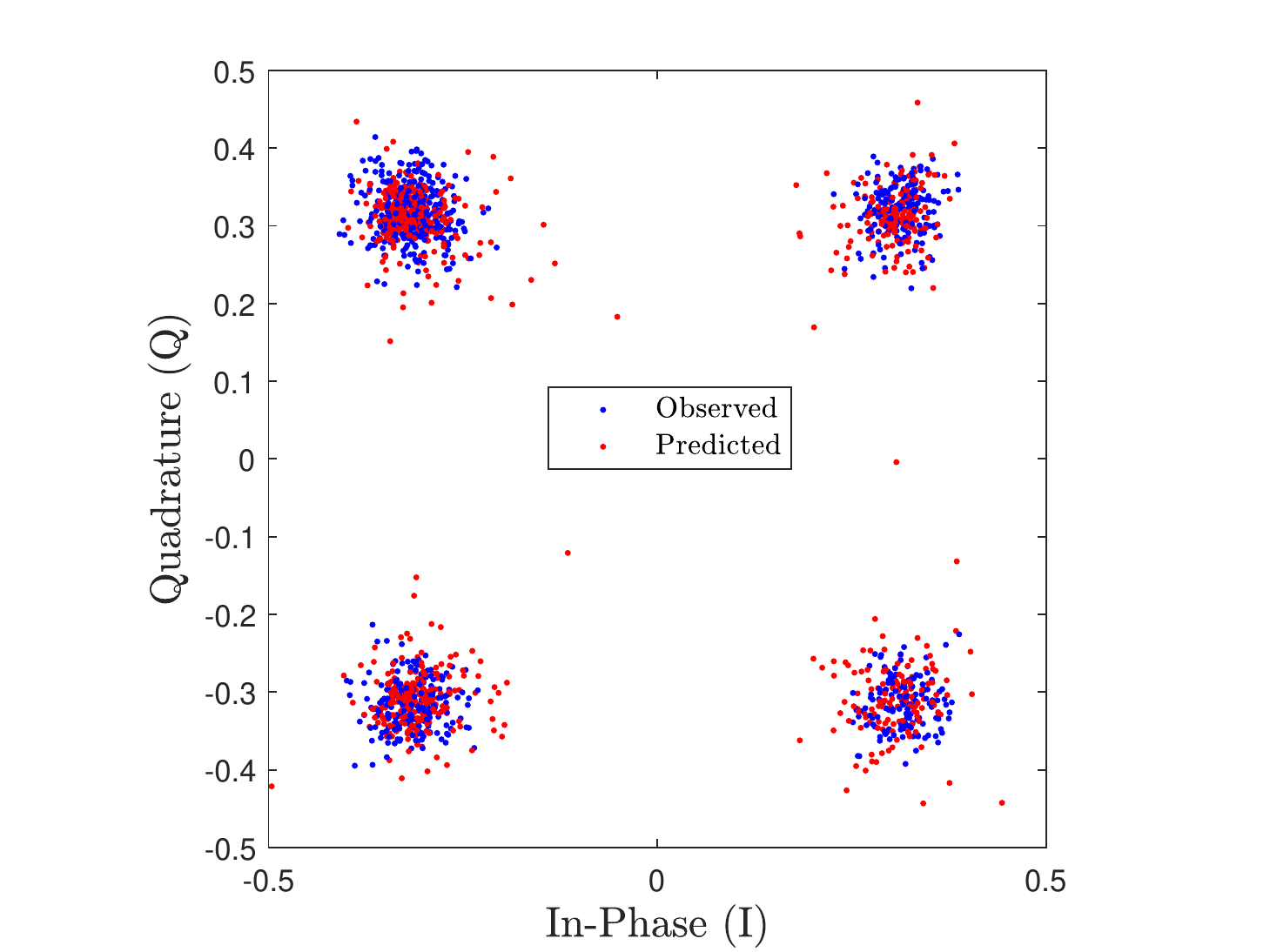}
        \\[-1.5mm]
        {\scriptsize (a)}
        \end{minipage}
        \begin{minipage}[b]{0.49\linewidth}
            \centering
            \includegraphics[width=4.9cm]{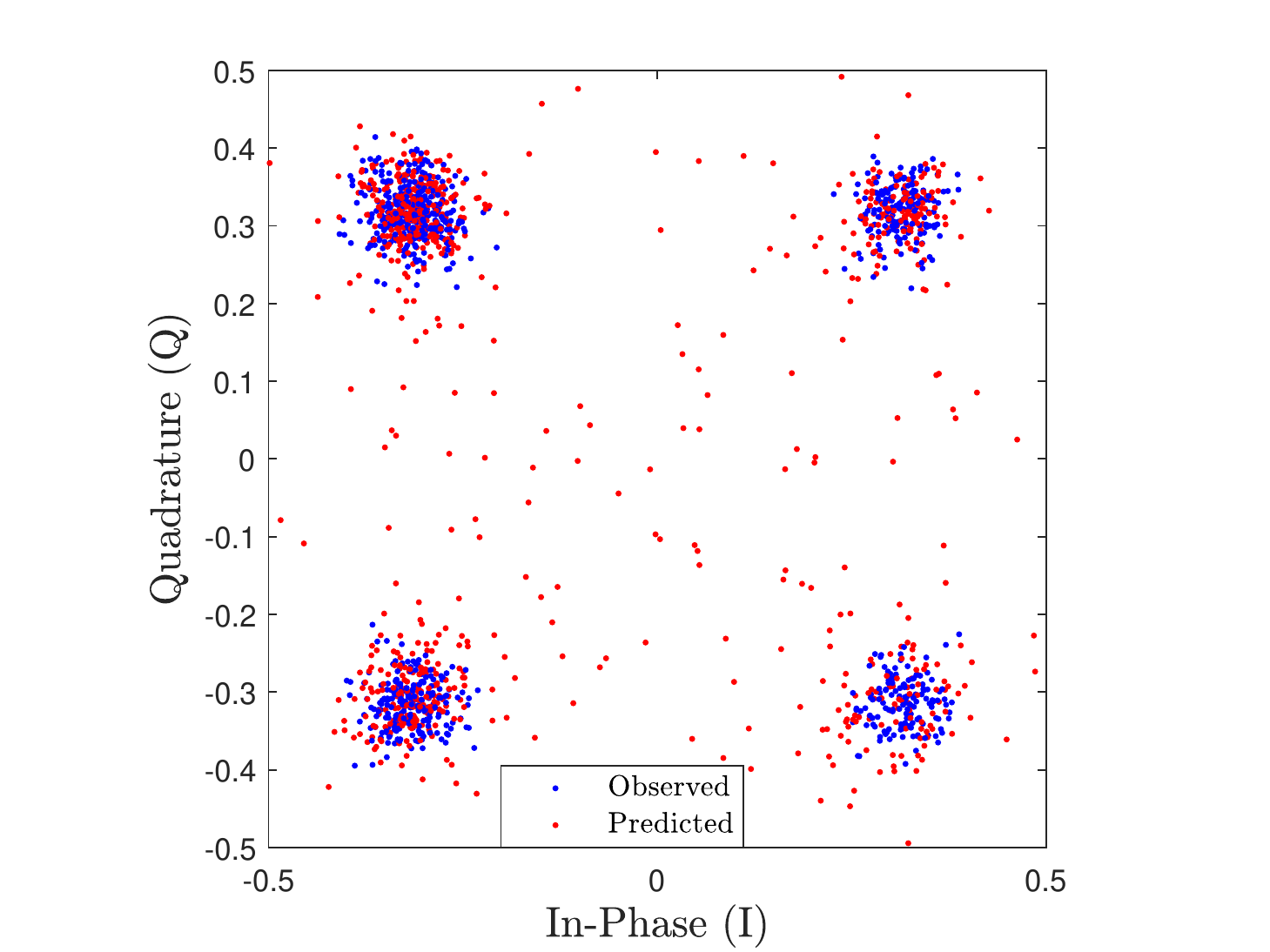}
            \\[-1.5mm]
            {\scriptsize (b)}
        \end{minipage}
        \begin{minipage}[b]{0.49\linewidth}
            \centering
            \includegraphics[width=4.9cm]{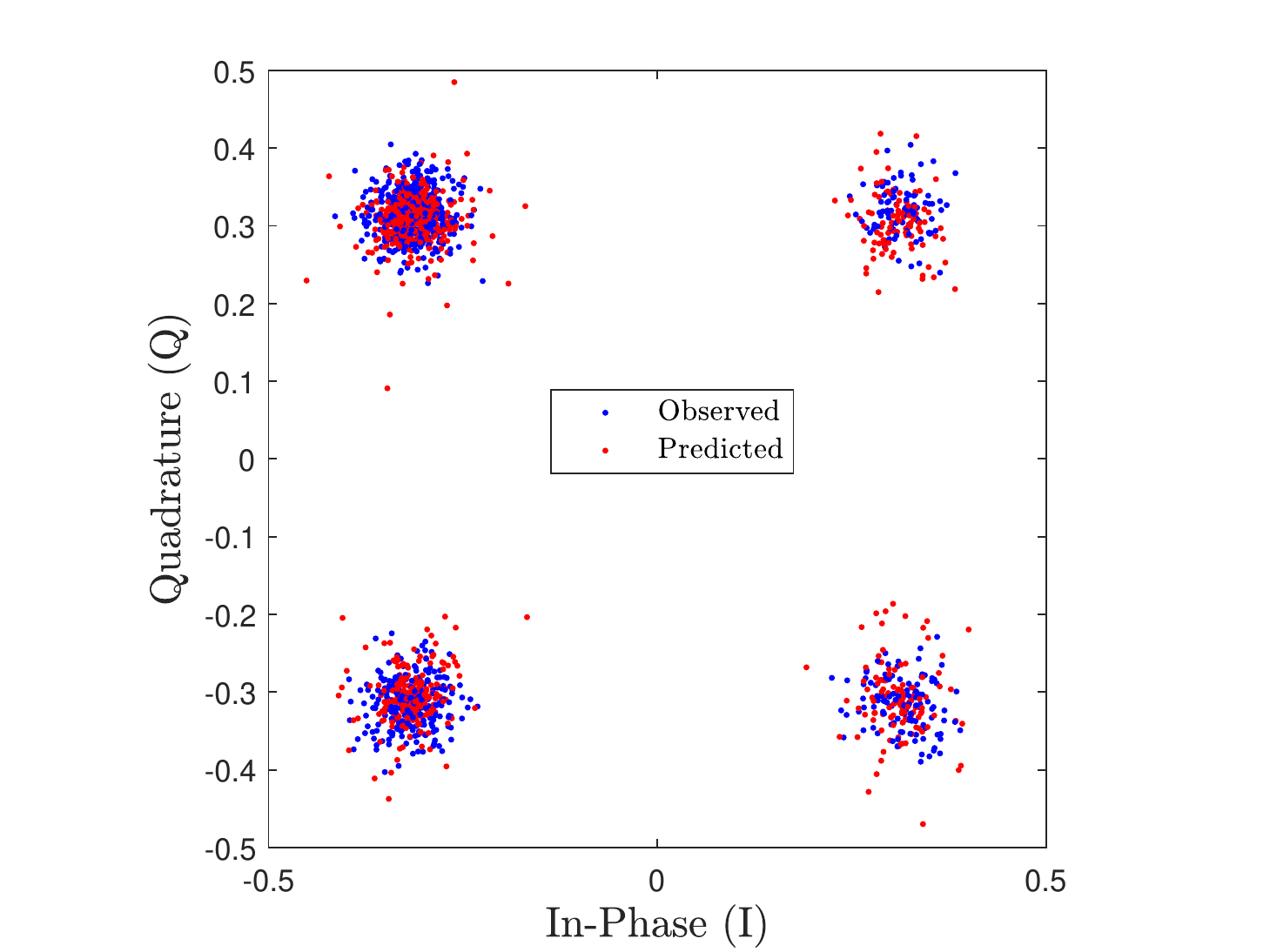}
            \\[-1.5mm]
            {\scriptsize (c)}
        \end{minipage}
        \begin{minipage}[b]{0.49\linewidth}
            \centering
            \includegraphics[width=4.9cm]{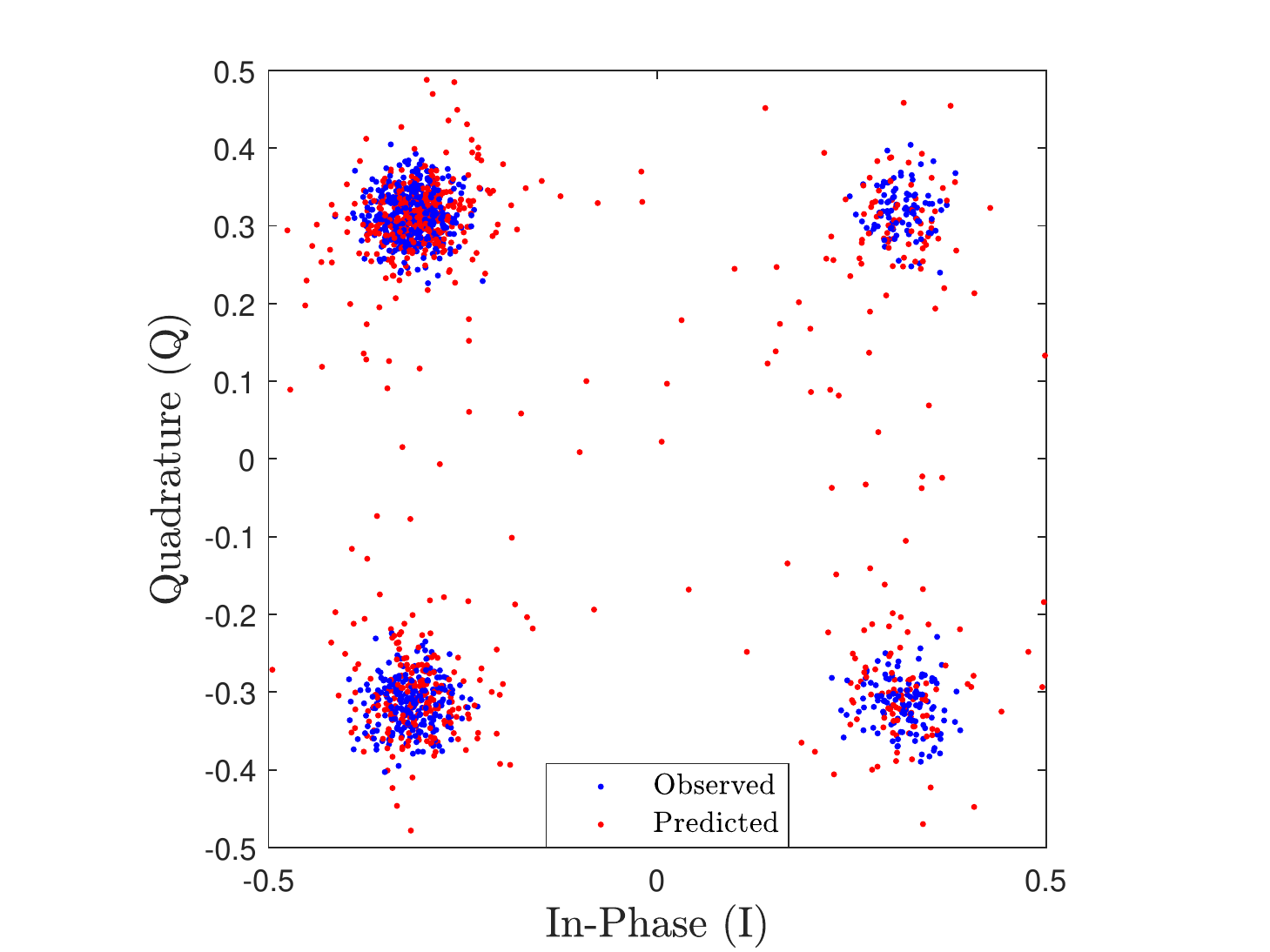}
            \\[-1.5mm]
            {\scriptsize (d)}
        \end{minipage}
        \caption{An example visualizing the predicted and observed RF signals transmitted by the 2 vehicles using different configurations. Predicted RF signal from: (a) vehicle 1 using $\mathrm{M_{1}}{=}5$, $\mathrm{M_{2}}{=}5$, (b) vehicle 1 using $\mathrm{M_{1}}{=}25$, $\mathrm{M_{2}}{=}25$, (c) vehicle 2 using $\mathrm{M_{1}}{=}5$, $\mathrm{M_{2}}{=}5$, (d) vehicle 2 using $\mathrm{M_{1}}{=}25$, $\mathrm{M_{2}}{=}25$.}
            \label{fig_situation1_PredictedRF}
    \end{center}
\end{figure}
\begin{figure}[t!]
    \begin{center}
        \begin{minipage}[b]{.49\linewidth}
        \centering
            \includegraphics[width=4.8cm]{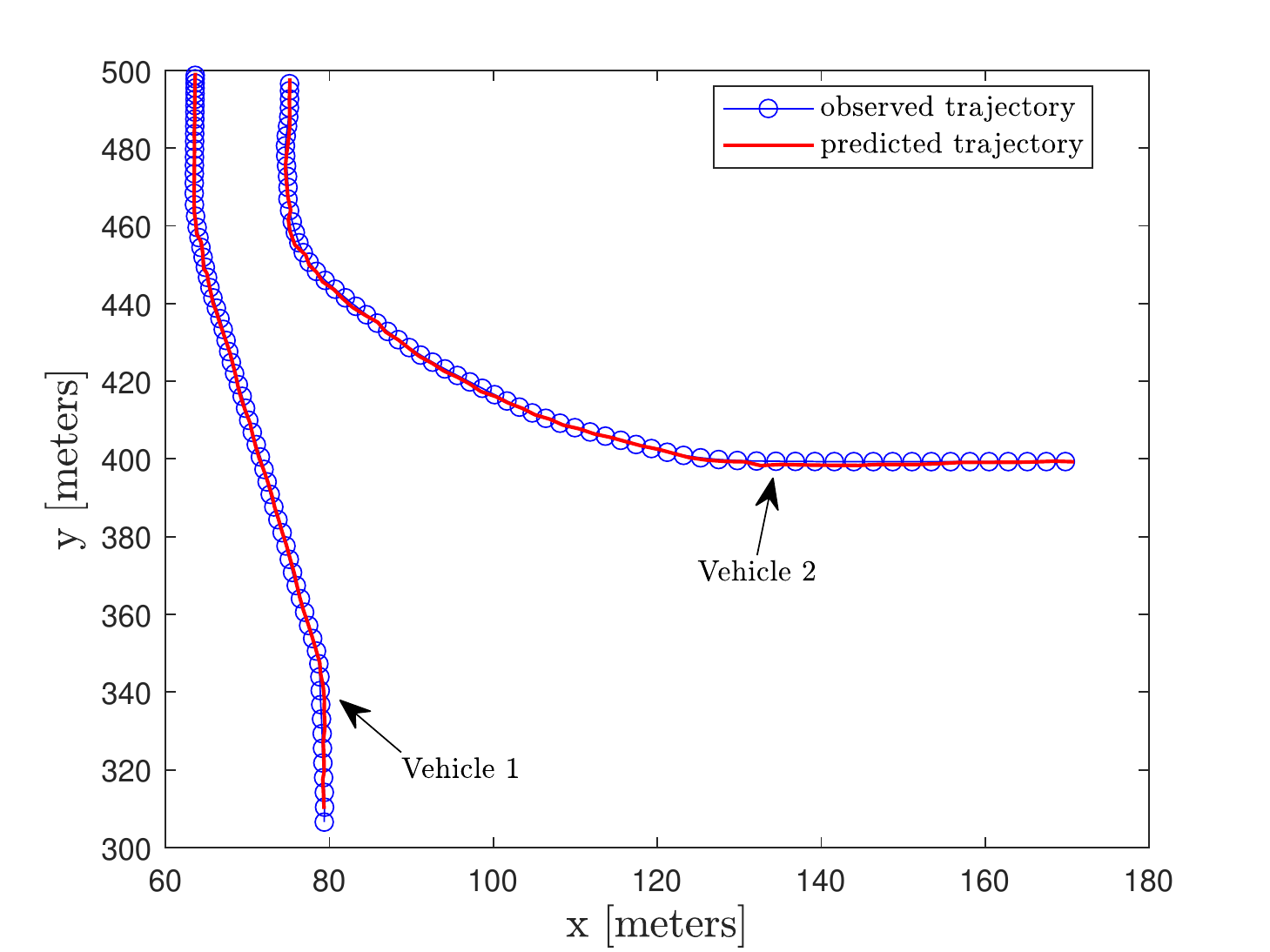}
        \\[-1.0mm]
        {\scriptsize (a)}
        \end{minipage}
        \begin{minipage}[b]{0.49\linewidth}
            \centering
            \includegraphics[width=4.8cm]{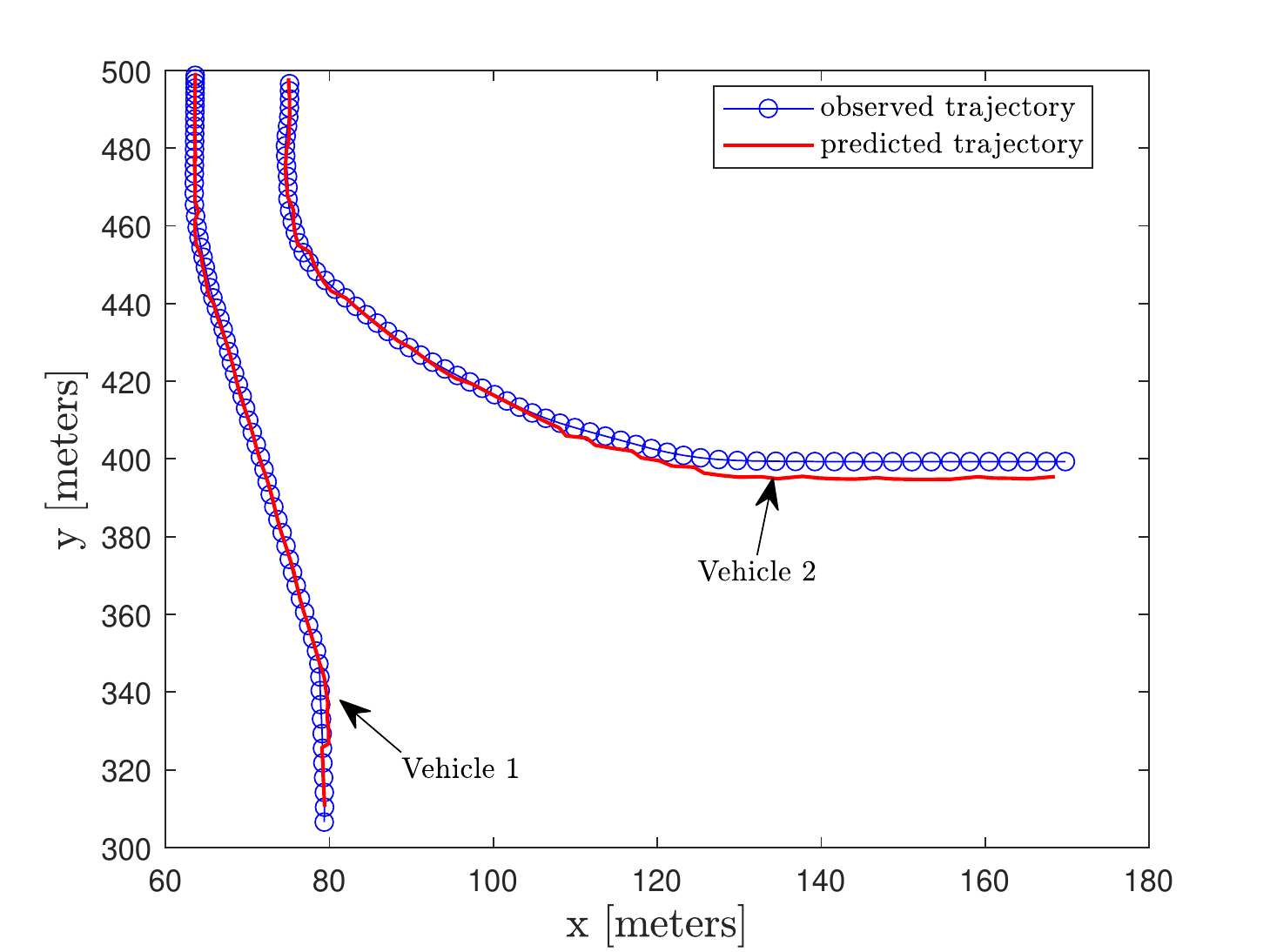}
            \\[-1.0mm]
            {\scriptsize (b)}
        \end{minipage}
        \caption{An example visualizing the predicted and observed trajectories of two vehicles interacting in the environment. (a) $\mathrm{M_{1}}{=}5$, $\mathrm{M_{2}}{=}5$, (b) $\mathrm{M_{1}}{=}25$, $\mathrm{M_{2}}{=}25$.}
            \label{fig_situation1_VehiclesTrajectories}
    \end{center}
\end{figure}
\begin{figure}[ht!]
    \begin{center}
        \begin{minipage}[b]{.49\linewidth}
        \centering
            \includegraphics[width=4.8cm]{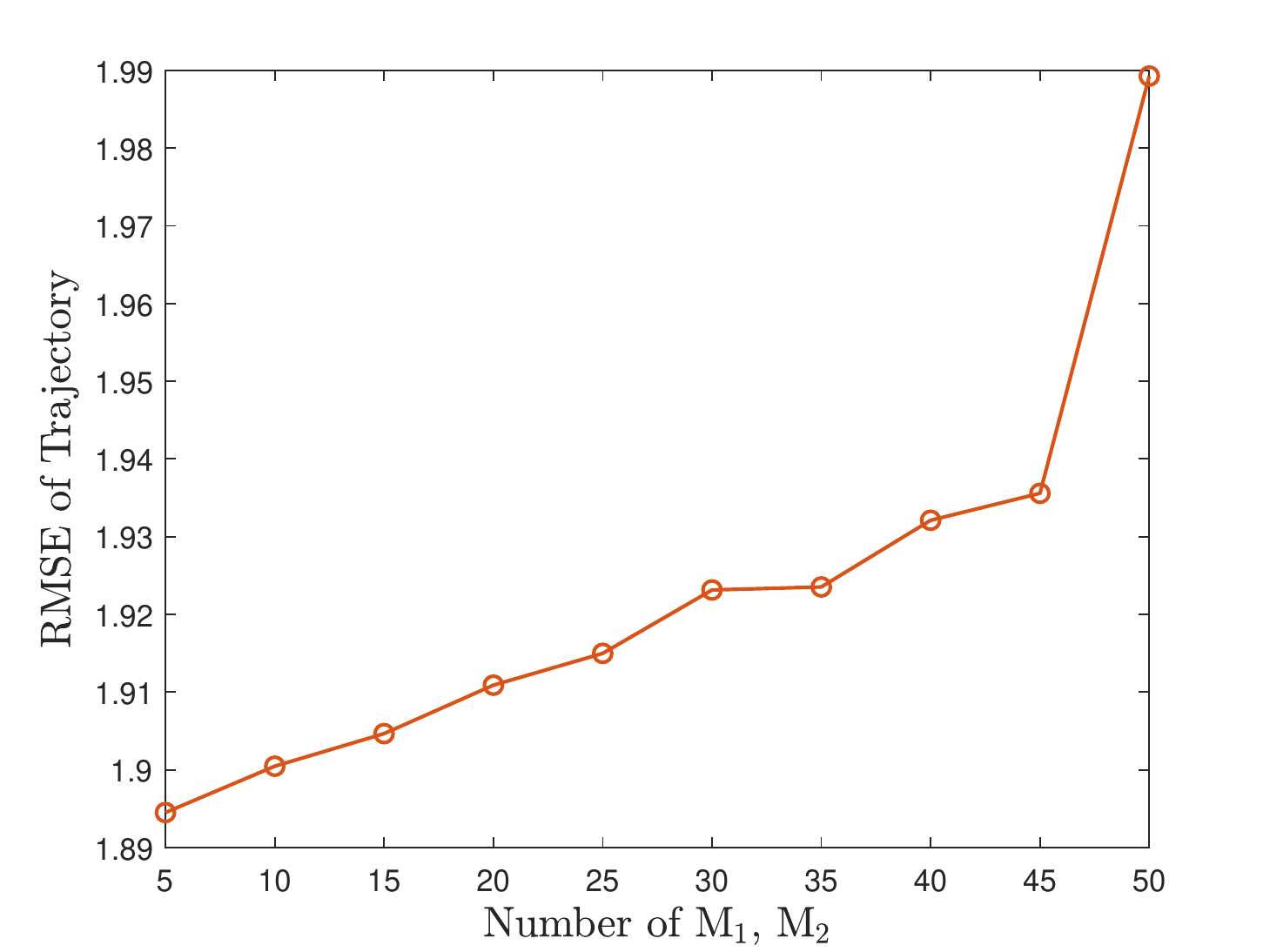}
        \\[-1.0mm]
        {\scriptsize (a)}
        \end{minipage}
        \begin{minipage}[b]{0.49\linewidth}
            \centering
            \includegraphics[width=4.8cm]{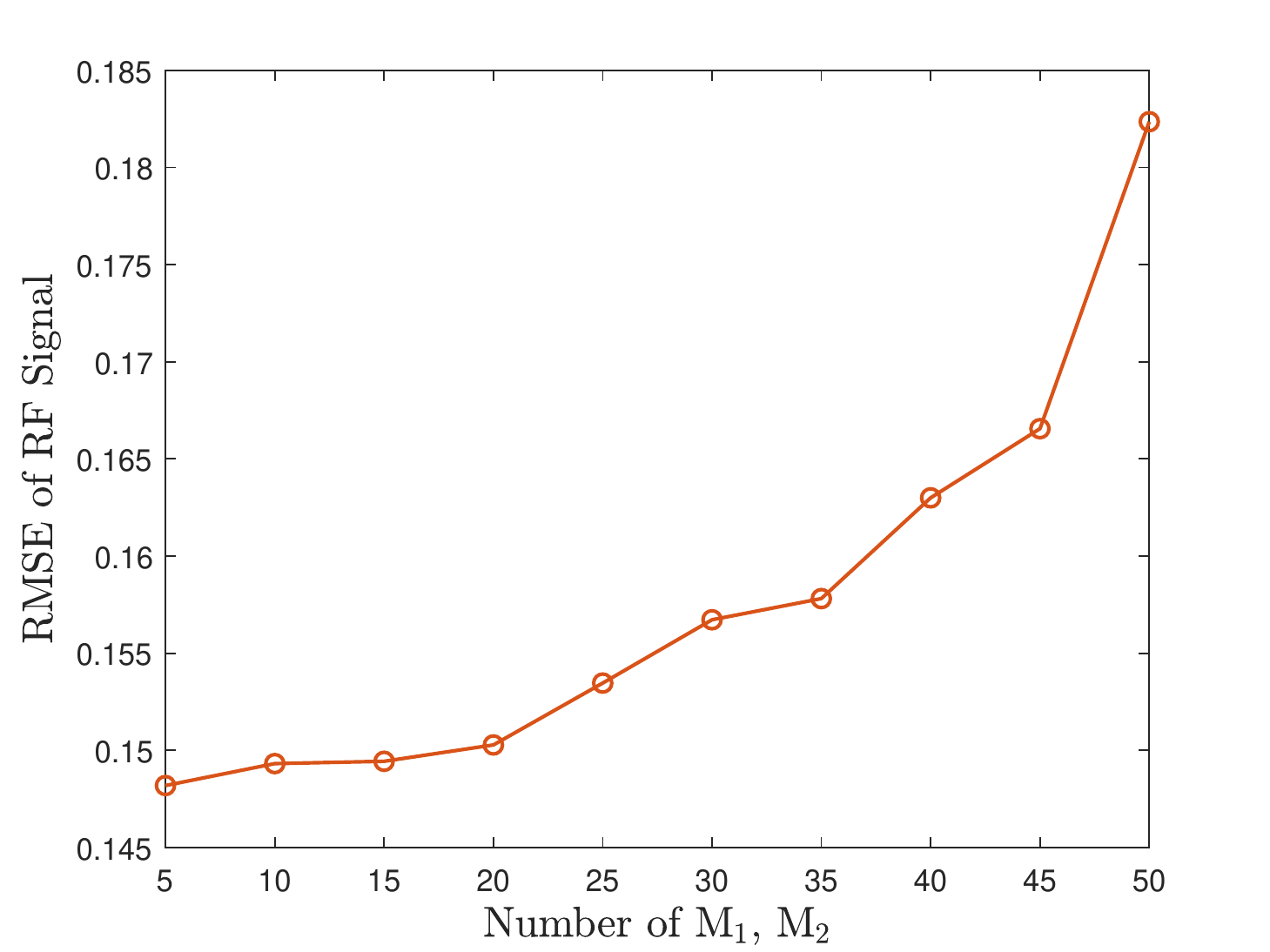}
            \\[-1.0mm]
            {\scriptsize (b)}
        \end{minipage}
        \caption{The average RMSE after testing different experiences and examples of: (a) trajectories and (b) RF signals.}
            \label{fig_rmse_onTraj_onSig}
    \end{center}
\end{figure}

Fig.~\ref{fig_situation1_PredictedRF} illustrates an example comparing between predicted RF signals and observed ones based on two different configurations in learning the interactive matrix (as shown in Fig.~\ref{fig_interactiveMatrices}). Also, Fig.~\ref{fig_situation1_VehiclesTrajectories} illustrates an example comparing between the predicted and observed trajectories of the two vehicles using the two interactive matrices depicted in Fig.~\ref{fig_interactiveMatrices}. From Fig.~\ref{fig_situation1_PredictedRF} and Fig.~\ref{fig_situation1_VehiclesTrajectories} we can see that using an interactive matrix with less clusters allows to perform better predictions compared to that with more clusters. This can be validated by observing Fig.~\ref{fig_rmse_onTraj_onSig} that illustrates the RMSE values versus different number of clusters related to the two models representing the dynamics of the received RF signals and the vehicles' trajectories. It can be seen that as the number of clusters increases the RMSE error increases, since adding more clusters decreases the firing probability that explains the possibility to be in one of the $M_{2}$ clusters of the second model conditioned in being in a certain cluster of the first model.

\begin{figure}[t!]
    \centering
    \includegraphics[width=6.5cm]{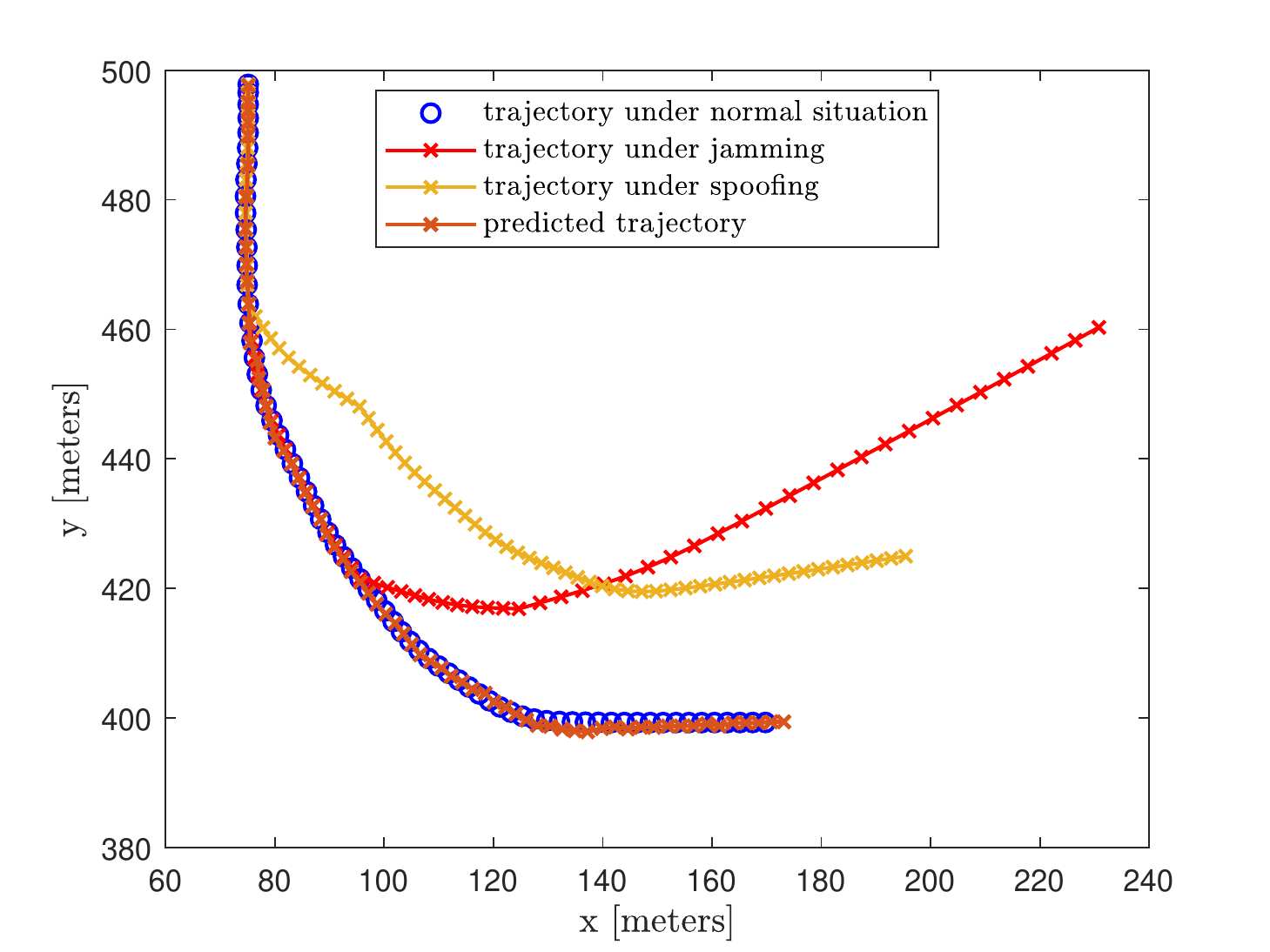}
    \caption{Vehicle's trajectory under: normal situation, jamming and spoofing.}
    \label{fig_exNormal_Spoofed_JammedTrajectories}
\end{figure}

Fig.~\ref{fig_exNormal_Spoofed_JammedTrajectories} illustrates an example of vehicle's trajectory under normal situation (i.e., jammer and spoofer are absent), under jamming attacks and under spoofing attacks. Also the figure shows the predicted trajectory which should follow the same dynamic rules learned during a normal situation. After that, we implemented the IM-MJPF on the learned C-GDBN to perform multiple predictions, i.e., to predict the RF signal that the RSU is expecting to receive from a certain vehicle and the corresponding trajectory that the vehicle is supposed to follow. IM-MJPF through the comparison between multiple predictions and observations, produces multiple abnormality signals as defined in \eqref{eq_CLA1} and \eqref{eq_CLA2} which are used to detect the jammer and the spoofer.

\begin{figure}[t!]
    \begin{center}
        \begin{minipage}[b]{.92\linewidth} 
        \centering
            \includegraphics[height=2.6cm]{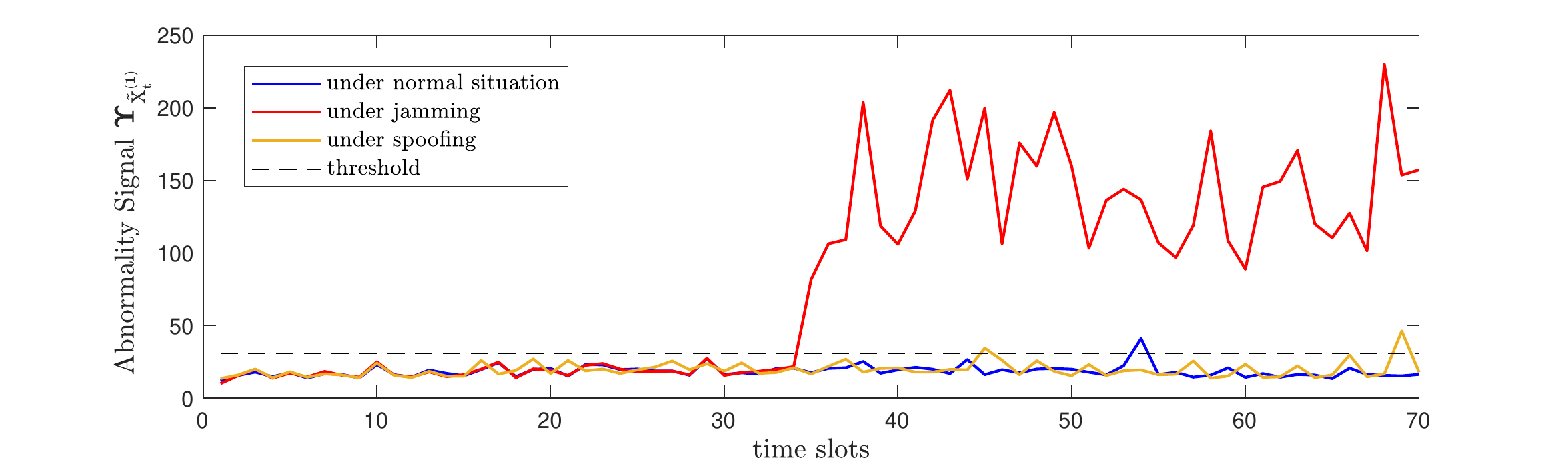}
        \\[-1.5mm]
        {\scriptsize (a)}
        \end{minipage}
        \begin{minipage}[b]{.92\linewidth} 
            \centering
            \includegraphics[height=2.6cm]{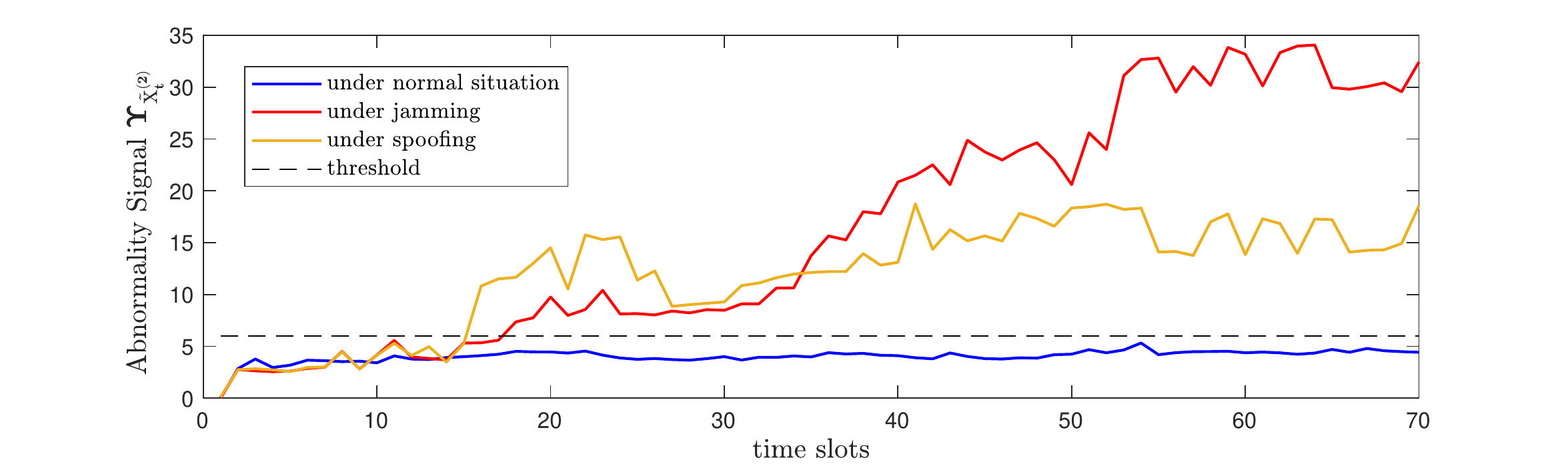}
            \\[-1.5mm]
            {\scriptsize (b)}
        \end{minipage}
        \caption{Abnormality Signals related to the example shown in Fig.\ref{fig_exNormal_Spoofed_JammedTrajectories}: (a) abnormality indicators related to the RF signal, (b) abnormality indicators related to the trajectory.}
            \label{fig_abnormalitySignals_JammerSpoofer}
    \end{center}
\end{figure}

Fig.~\ref{fig_abnormalitySignals_JammerSpoofer} illustrates the multiple abnormality signals related to the example shown in Fig.~\ref{fig_exNormal_Spoofed_JammedTrajectories}. We can observe that the abnormal signals related to both RF signal (Fig.~\ref{fig_abnormalitySignals_JammerSpoofer}-(a)) and trajectory (Fig.~\ref{fig_abnormalitySignals_JammerSpoofer}-(b)) are below the threshold under normal situations. This proves that RSU learned the correct dynamic rules of how RF signals and trajectories evolve when the jammer and spoofer are absent (i.e., under normal situations). Also, we can see that the RSU can notice a high deviation on both the RF signal and the corresponding trajectory due to a jamming interference from what it has learned so far by relying on the abnormality signals. In contrast, we can see that under spoofing attacks, RSU notice a deviation only on the trajectory and not on the RF signal since the spoofer has affected only the positions without manipulating the RF signal. In addition, it is obvious how the proposed method allows the RSU to identify the type of abnormality occurring and to explain the cause of the detected abnormality (i.e., understanding if it was because of a jammer attacking the V2I link or a spoofer attacking the satellite link).
\begin{figure}[t!]
    \centering
    \includegraphics[height=3.2cm]{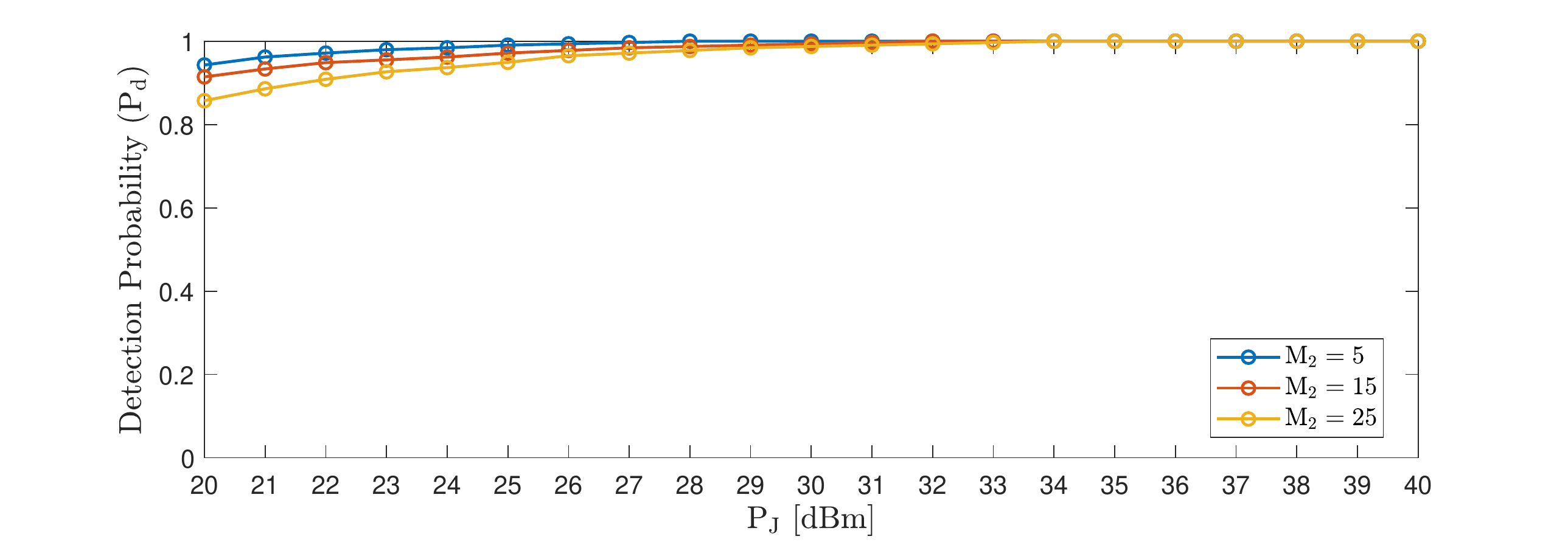}
    \caption{Detection probability ($\mathrm{P_{d}}$) versus jammer's power ($\mathrm{P_{J}}$) using different number of clusters $\mathrm{M}_{2}$.}
    \label{fig_jammerDetectionProb}
\end{figure}
\begin{figure}[t!]
    \centering
    \includegraphics[height=3.2cm]{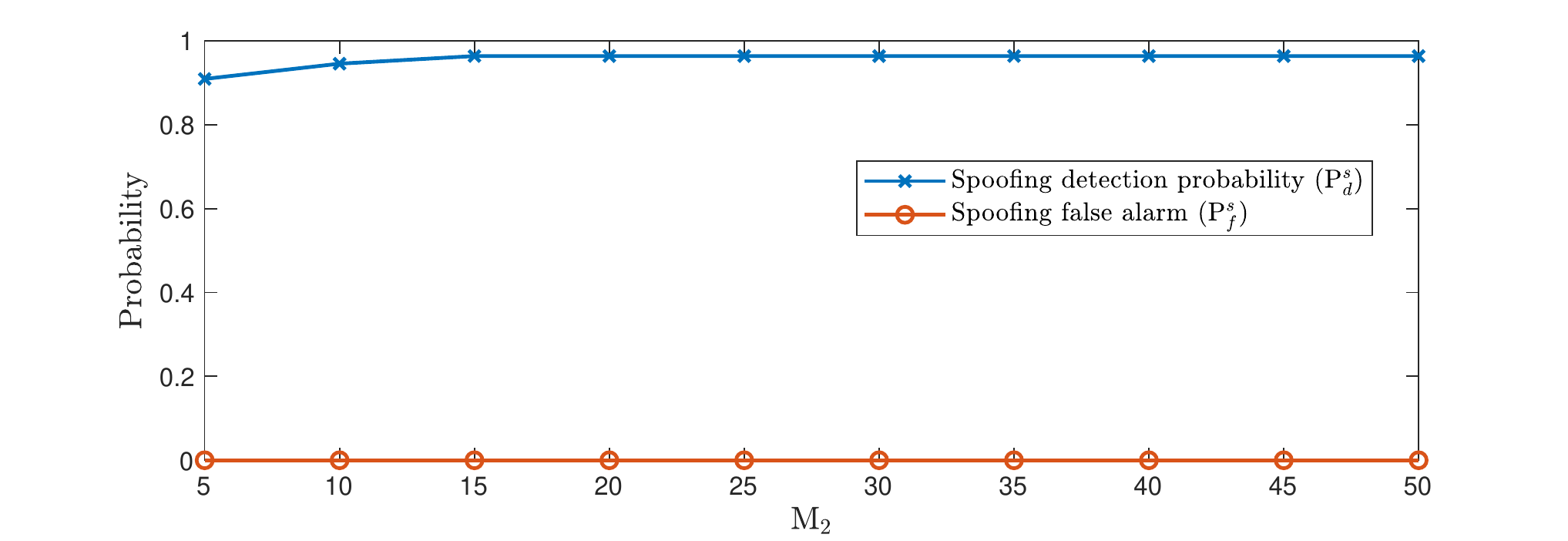}
    \caption{Spoofing detection probability ($\mathrm{P}_{d}^{s}$) and spoofing false alarm ($\mathrm{P}_{f}^{s}$) versus the number of clusters $\mathrm{M}_{2}$.}
    \label{fig_spooferDetectionProb}
\end{figure}

Fig.~\ref{fig_jammerDetectionProb} shows the overall performance of the proposed method in detecting the jammer by testing many situations and examples and by considering different jamming powers which ranges from $20$dBm to $40$dBm. It can be seen that the proposed method is able to detect the jammer with high probabilities (near $1$) and by considering low and high jamming powers. Also, the figure compares the performance in detecting the jammer by varying the number of clusters ($M_{2}$).
Fig.~\ref{fig_spooferDetectionProb} shows the overall performance of the proposed method in detecting the spoofer by testing different different examples of driving maneuvers. It can be seen that the RSU is able to detect the spoofer with high detection probability and null false alarm versus different number of clusters.

\section{Conclusion}
A joint detection method of GPS spoofing and jamming attacks is proposed. The method is based on learning a dynamic interactive model encoding the cross-correlation between the received RF signals from multiple vehicles and their corresponding trajectories. Simulation results show the high effectiveness of the proposed approach in jointly detecting the GPS spoofer and jammer attacks. 
Subsequent work will extend the system model to consider more than two vehicles with different channel conditions and various modulation schemes to evaluate the effectiveness of the proposed method.

\bibliographystyle{IEEEtran}
\bibliography{Reference}

\vspace{12pt}

\end{document}